\documentclass[acmlarge,screen]{acmart}
\usepackage{graphicx}
\usepackage{amsmath,amsfonts}
\usepackage[normalem]{ulem}
\usepackage{subfigure}
\usepackage{flushend}
\usepackage{latexsym}
\usepackage{paralist}
\usepackage{xspace}
\usepackage{multirow}
\usepackage{dsfont}
\usepackage{multicol}
\usepackage{etoolbox}
\usepackage{courier}
\usepackage{booktabs}
\usepackage{setspace}
\usepackage{amscd}
\usepackage{relsize}
\usepackage{colortbl}
\usepackage{listings}
\usepackage{enumitem}
\usepackage{soul}
\usepackage{cancel}
\usepackage{textcase}
\usepackage{dblfloatfix}
\usepackage[flushleft]{threeparttable}
\usepackage{makecell}
\usepackage{etoolbox}
\usepackage{caption}  
\usepackage{booktabs}
\usepackage{float} 
\usepackage{tikz-cd}
\usepackage[most]{tcolorbox} 
\usepackage{sistyle}
\SIthousandsep{,}
\usepackage{rotating}
\usepackage{algpseudocode}
\usepackage{multirow}
\usepackage{tikz}
\usepackage{booktabs}
\usepackage{threeparttable}
\usepackage{pifont}
\usepackage{tabularx} 
\usepackage{booktabs}
\makeatletter
\newcommand{\blackmailto}[1]{\href{mailto:#1}{\textcolor{black}{#1}}}
\authorsaddresses{%
Authors' Contact Information: \href{https://orcid.org/0009-0004-3200-4871}{Guosheng Wang}, Department of Computing, The Hong Kong Polytechnic University, China,\blackmailto{river.wang@connect.polyu.hk}; \href{https://orcid.org/0009-0000-1543-3819}{Shen Wang}, Department of Computing, The Hong Kong Polytechnic University, China, \blackmailto{wangshen@tagsys.org}; \href{https://orcid.org/0000-0001-9651-0811}{Lei Yang}~(corresponding author), Department of Computing, and Shenzhen Research Institute, The Hong Kong Polytechnic University, China, \blackmailto{young@tagsys.org}.}
\makeatother

\def\ie{i.e.,\xspace}

\DeclareMathOperator*{\argmin}{argmin\,}

\newcommand{\oursystem}{{LocGPT+}\xspace}
\newcommand{\nerf}{NeRF$^2$\xspace}
\newcommand{\rfnerf}{\textsf{RF-NeRF}\xspace}
\newcommand{\llm}{\textsf{LM}\xspace}
\newcommand{\rfrp}{\textsf{RFGP}\xspace}

\graphicspath{{oldfigs/}{figs/}}

\begin{document}
\setcopyright{cc}
\setcctype{by}
\acmJournal{IMWUT}
\acmYear{2026} \acmVolume{10} \acmNumber{3} \acmArticle{156}
\acmMonth{9} \acmDOI{10.1145/3831969}

\title[Radiance-Field Guided Pretraining: Scaling Localization Models with Unlabeled Wireless Signals]{Radiance-Field Guided Pretraining: \\ Scaling Localization Models with Unlabeled Wireless Signals}

\author{Guosheng Wang}
\orcid{0009-0004-3200-4871}
 \affiliation{%
  \institution{The Hong Kong Polytechnic University}
  \country{China}
}
\email{river.wang@connect.polyu.hk}

\author{Shen Wang}
\orcid{0009-0000-1543-3819}
 \affiliation{%
  \institution{The Hong Kong Polytechnic University}
  \country{China}
}
\email{wangshen@tagsys.org}

\author{Lei Yang}
\orcid{0000-0001-9651-0811}
\authornote{Corresponding author.}
 \affiliation{%
  \institution{The Hong Kong Polytechnic University}
  \country{China}
}
\email{young@tagsys.org}

\renewcommand{\shortauthors}{Wang et al.}

\begin{abstract}
	Radio frequency (RF)-based indoor localization offers significant promise for applications such as indoor navigation, augmented reality, and pervasive computing. While deep learning has greatly enhanced localization accuracy and robustness, existing models still struggle with cross-scene generalization because they rely on scene-specific labeled data. To address this, we introduce Radiance-Field Guided Pretraining (RFGP), a novel self-supervised pretraining framework. Unlike conventional pretraining approaches, RFGP utilizes a physics-guided asymmetric autoencoder architecture. Specifically, a large localization model (LM) encodes received RF spectra into latent representations, while a neural radio-frequency radiance field (RF-NeRF) acts as a physics-informed decoder to reconstruct the original spectra via ray tracing. This unique design encourages the LM to learn more transferable representations from unlabeled RF data, which can be collected continuously with minimal effort. To evaluate RFGP, we constructed a massive-scale dataset comprising 7,327,321 RF samples across 100 diverse scenes using four wireless technologies (RFID, BLE, WiFi, and IIoT). Extensive evaluations on unseen test scenes show that the RFGP-pretrained LM reduces localization error by over 40\% compared to training from scratch and by 21\% compared to standard supervised pretraining, underscoring the potential of RFGP as a scalable pretraining framework for generalized indoor localization. The dataset and implementation code are available in our \href{https://github.com/moonriver0922/RFGP}{repository}\footnote{\url{https://github.com/moonriver0922/RFGP}}.
\end{abstract}


\maketitle

\section{INTRODUCTION}
\label{section:introduction}
 \begin{figure}[!t]
 	\centering
\includegraphics[width=0.8\linewidth]{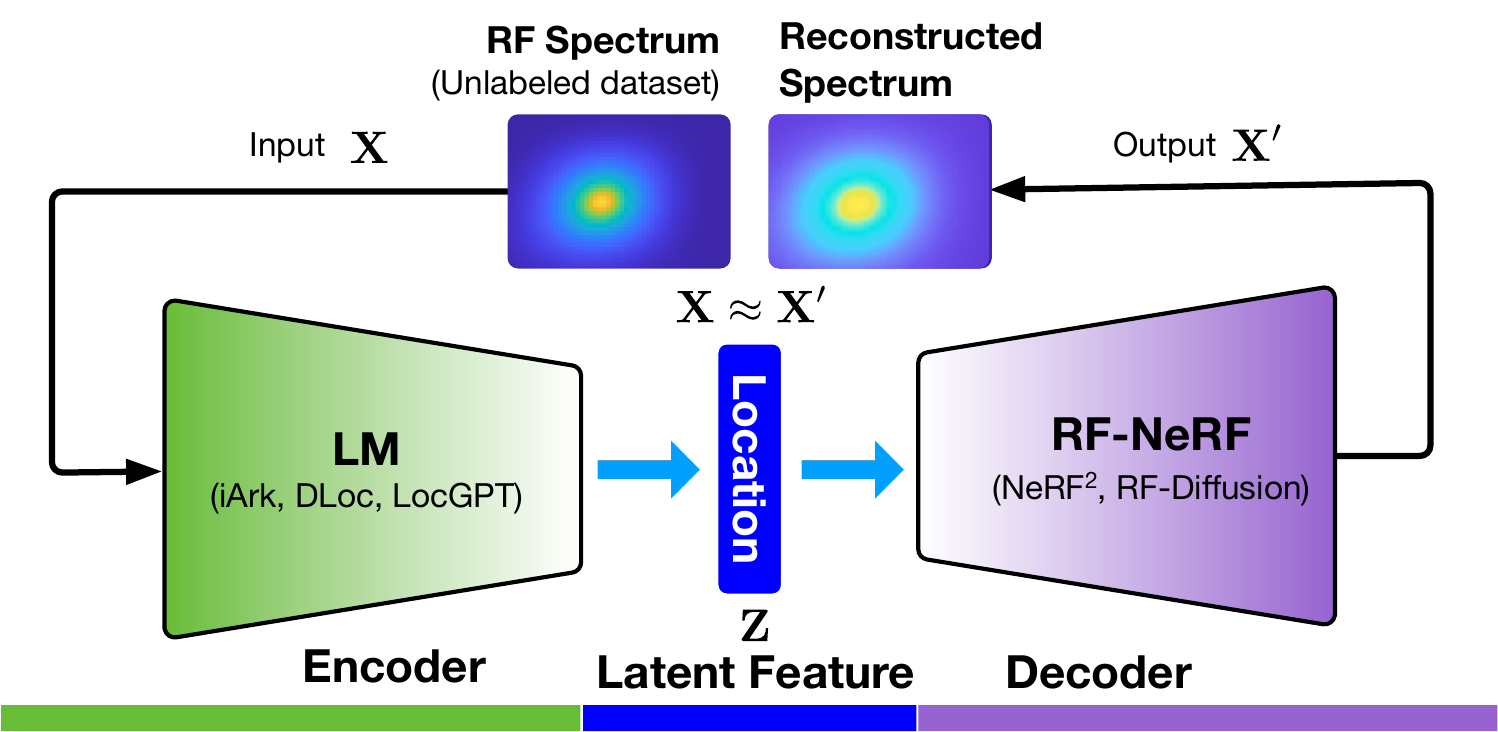}
	\caption{Radiance-Field Guided Pretraining for Large Localization Models. \textnormal{This approach integrates an LM and an RF-NeRF model into a unified encoder-decoder training framework, designed to pretrain the LM for extracting generalizable features pertinent to localization.}}	
	\label{fig:architecture}
 \end{figure}
Radio frequency (RF)-based indoor localization estimates device positions by analyzing wireless signals received at base stations, enabling reliable tracking in environments where GPS is unavailable or unreliable. High-precision indoor localization supports a wide range of applications, including indoor navigation, augmented reality, location-aware pervasive computing, targeted advertising, and social networking. Consequently, the task of tracking IoT devices within built environments has become a growing area of commercial and academic interest, giving rise to a substantial body of research over the past two decades~\cite{yang2014tagoram, ma2017minding, xie2019md, xie2018swan, adib20133d, adib2014multi, zhao2018rf, zhao2018through, ma2014accurate, hui2019radio, haniz2017novel, youssef2005horus, sen2012you, yang2012locating, liu2012push, wang2012no, ni2004landmarc, 10.1145/2486001.2486029, pan2008transfer}. 

In the wake of the deep learning surge, recent studies~\cite{ayyalasomayajula2020deep, an2020general, li2021deep, qian2021supervised, zhan2021deepmtl, zhao2024understanding} have demonstrated the transformative potential of deep learning-based localization models (LMs) over traditional algorithms, particularly in addressing challenges related to accuracy, robustness, and adaptability. These models reframe indoor localization as an optimization task, using radio-frequency signals received at base stations to probabilistically infer the spatial coordinates of RF devices~\cite{zhao2024understanding, yang2014tagoram}. This paradigm aligns naturally with the strengths of deep neural networks (DNNs), which excel at uncovering intricate patterns in high-dimensional data. 

However, a key challenge persists: existing models exhibit poor cross-scene generalization, with performance closely tied to the spatial layouts and RF characteristics of their training environments. Models trained on a single scene often degrade significantly when applied to unseen or dynamically changing settings, primarily because they lack the ability to extract and transfer invariant features across heterogeneous scenarios. To address this, recent work such as LocGPT~\cite{zhao2024understanding} has explored pretraining large localization models (LMs) on aggregated multi-scene datasets. Yet this pretraining method relies heavily on supervised learning, creating a major bottleneck: the need for extensive, high-quality position labels. 

{\color{black}
To address the data-hungry nature of LMs, recent pioneering work has begun exploring self-supervised pretraining using unlabeled wireless signals. However, most existing approaches rely on conventional symmetric masked reconstruction, which treats RF signals merely as numerical sequences and fails to capture their underlying physical propagation laws. 

\textcolor{black}{In this work, we propose Radiance-Field Guided Pretraining (\rfrp), a novel self-supervised pretraining framework. Unlike prior works that focus solely on data scaling, the core innovation of \rfrp lies in its asymmetric architecture, which uses a physical prior as the decoder. As illustrated in Fig.~\ref{fig:architecture}, \rfrp couples a large localization model (encoder) with a neural radio-frequency radiance field (RF-NeRF) acting as a physics-informed decoder. The LM takes in the spectrum received from an RF base station and outputs a latent representation. This latent feature is then utilized by the RF-NeRF decoder, which reconstructs the spectrum through a ray-tracing-inspired rendering process. By introducing a physics-guided reconstruction objective, \rfrp encourages the encoder to learn representations that are less tied to scene-specific spectral artifacts.}
}

$\bullet$ \emph{What scene-agnostic features should the LM extract?}  Regardless of whether triangulation or trilateration is used, the core challenge in localization lies in accurately identifying the line-of-sight (LoS) path, which directly reflects the geometric relationship between the transmitter and base stations and is independent of the scene layout. However, multipath effects—caused by RF signals reflecting off surrounding objects—result in received signals being a superposition of multiple propagation paths, making the spectra highly scene-dependent. Thus, a desirable localization representation should be sensitive to geometric cues associated with LoS propagation while being robust to scene-dependent multipath distortions.

To this end, we extend the classical Transformer encoder into a 570-million-parameter localization model, referring to \oursystem, by integrating a key enhancement: the Mixture of Experts (MoE) architecture. MoE is particularly well-suited for localization tasks due to its ability to handle complex, diverse spatial configurations. We hypothesize that each expert learns to specialize in different environment types, such as open areas or cluttered indoor settings. As the model is exposed to a wider range of scenes during training, the ensemble of experts collectively improves localization performance.
Critically, the MoE framework enhances generalization by dynamically routing each input to the most relevant experts based on learned similarity. When encountering a new scene, the model can leverage knowledge from previously seen environments with similar characteristics. By distributing learning across multiple experts and selectively activating the most relevant ones, the MoE architecture significantly improves the localization model's generalization.


$\bullet$ \emph{How does the RF-NeRF guide the pretraining of the LM?} RF-NeRFs are neural-network-based models originally inspired by NeRF~\cite{mildenhall2020nerf}, designed to estimate radiance fields representing how radio frequency signals propagate within spatial regions. Recent works such as NeRF$^2$~\cite{zhao2023nerf2} and RF-Diffusion~\cite{rfdiffusion} leverage RF-NeRF architectures to accurately simulate RF signal propagation using geometric ray-tracing principles.

\textcolor{black}{Since the RF-NeRF decoder reconstructs spectra through a ray-tracing-based formulation, it provides an inductive bias toward geometry-consistent representations. We hypothesize that this objective makes the encoder more sensitive to stable propagation cues, such as directionality and LoS-related structure, while reducing reliance on scene-specific multipath artifacts. Our ablation study in §9.2 provides empirical evidence that this ray-tracing-based prior improves downstream localization compared with a decoder of similar capacity that does not use ray tracing.}


\textbf{Contributions}. The core innovation of our work is the proposed RF-NeRF-guided asymmetric pretraining architecture. Rather than presenting unlabeled pretraining itself as entirely new, \oursystem shifts the paradigm by introducing a physical prior (the RF-NeRF decoder) into the self-supervised learning loop. \textcolor{black}{This design provides a physics-guided inductive bias that encourages the large localization model to learn more transferable representations. By leveraging abundant unlabeled multi-technology data (e.g., Wi-Fi, BLE, RFID, IIoT) across 100 scenes, our framework provides a highly scalable, cost-effective solution for training general-purpose, cross-scene localization models.}

\section{RELATED WORK} 
\label{section:related-work}

\subsection{Conventional Algorithms for Localization}
RF localization has been the focus of research, as detailed in studies~\cite{ni2003landmarc,yang2014tagoram,10.1145/3699777,10.1145/3729498}. RF signal metrics are used for localization, including: RSSI~\cite{ni2003landmarc}, carrier phase~\cite{ma2017drone,yang2014tagoram}, CSI~\cite{yang2013rssi,xie2019md}, Time-of-Flight (ToF)~\cite{mariakakis2014sail}, and AoA~\cite{an2020general,ayyalasomayajula2018bloc,kotaru2015spotfi,xiong2013arraytrack,gjengset2014phaser,xie2018swan,ayyalasomayajula2020deep}. This study focuses on AoA-based techniques, which encounter difficulties due to environmental factors such as complex multipath effects and non-line-of-sight (NLoS) situations. Previous work has tackled these challenges by identifying the direct path and reducing environmental impacts~\cite{xie2019md,ayyalasomayajula2018bloc}. However, these methods falter when multipath signals are closely spaced or when the direct path is completely blocked.

{\color{black}
\subsection{Deep Learning and Self-Supervised Localization}Recent research has increasingly leveraged deep learning models to enhance localization accuracy in complex indoor environments~\cite{ayyalasomayajula2020deep, an2020general, zheng2019zero, li2021train, zhao2024understanding,10.1145/3770667}. Concurrently, a variety of localization datasets have emerged to support robust learning-based systems~\cite{raza2019dataset, zheng2019zero, an2020general, dichasus2021, 10.1145/3659597}. To mitigate the high cost of collecting labeled data, recent studies have pioneered self-supervised pretraining for wireless localization. For instance, Radio Foundation Models~\cite{ott2024radiofoundationmodelspretraining} proposed a self-supervised masked reconstruction objective over unlabeled 5G channel measurements, while SelfLoc~\cite{electronics14132675} introduced robust self-supervised paradigms for IEEE 802.11az Wi-Fi. In this work, we introduce a newly designed model, \oursystem, inspired by LocGPT~\cite{zhao2024understanding}, that differs in two key architectural aspects. First, while LocGPT adopts a full Transformer architecture to model temporal correlations, \oursystem uses only the Transformer encoder to focus strictly on spatial correlations. Second, \oursystem incorporates a Mixture-of-Experts (MoE) architecture, expanding the model capacity to 570 million parameters to accommodate diverse environments. \textcolor{black}{Furthermore, while prior works~\cite{ott2024radiofoundationmodelspretraining, electronics14132675} rely on conventional self-supervised objectives, such as masked reconstruction or contrastive learning, our proposed pretraining framework, RFGP, distinguishes itself by employing a physics-guided asymmetric design to learn scene-agnostic geometric features from large-scale unlabeled data.}}\\
\subsection{Radio Frequency Radiance Fields}Inspired by neural rendering in computer vision, recent works have successfully modeled wireless signal propagation using implicit neural representations. WiNERT~\cite{orekondy2023winert} introduced neural ray tracing for differentiable wireless channel modeling, and NeRF$^2$~\cite{zhao2023nerf2,10.1145/3678572} proposed a neural RF radiance field framework for localization and 5G MIMO using turbo-learning. To address scalability limitations, subsequent works proposed training-free active sampling~\cite{gau2024active}, 3D Gaussian splatting for efficient reconstruction (WRF-GS~\cite{wen2024wrf}), dynamic environment adaptation (NeWRF~\cite{lu2024newrfdeeplearningframework}), and generative models for sparse-data scenarios (RF-Diffusion~\cite{rfdiffusion}). More recently, Yang et al.~\cite{yang2025generalizableradiofrequencyradiancefields} explored generalizable RF radiance fields for cross-scene spatial spectrum synthesis. \textcolor{black}{Our work builds upon these advancements but fundamentally shifts the application paradigm. Rather than using RF-NeRF directly as an end-to-end localizer or channel simulator, our framework uniquely repurposes it as a physics-informed decoder. This encourages the large localization model to learn representations that are more consistent with geometric propagation cues during self-supervised pretraining.}

\section{BACKGROUND}

\begin{figure}[t!]
    \centering
    \subfigure[Local Coordinate Systems]{
        \includegraphics[width=0.4\textwidth]{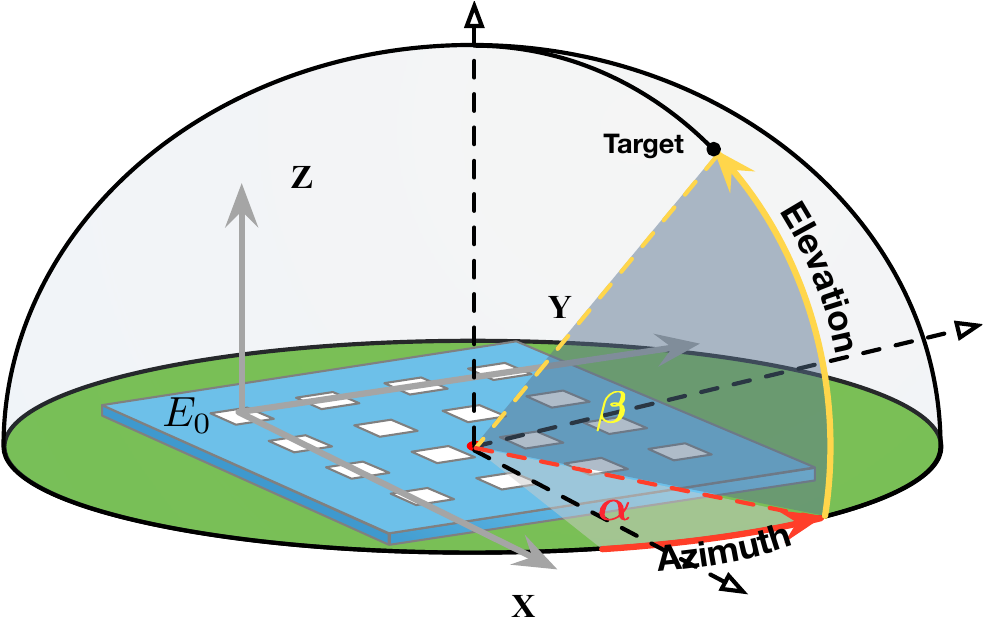}
        \label{fig:ant-array}
    }%
    \hspace{1.5cm}
    \subfigure[Triangulation]{
        \includegraphics[width=0.42\textwidth]{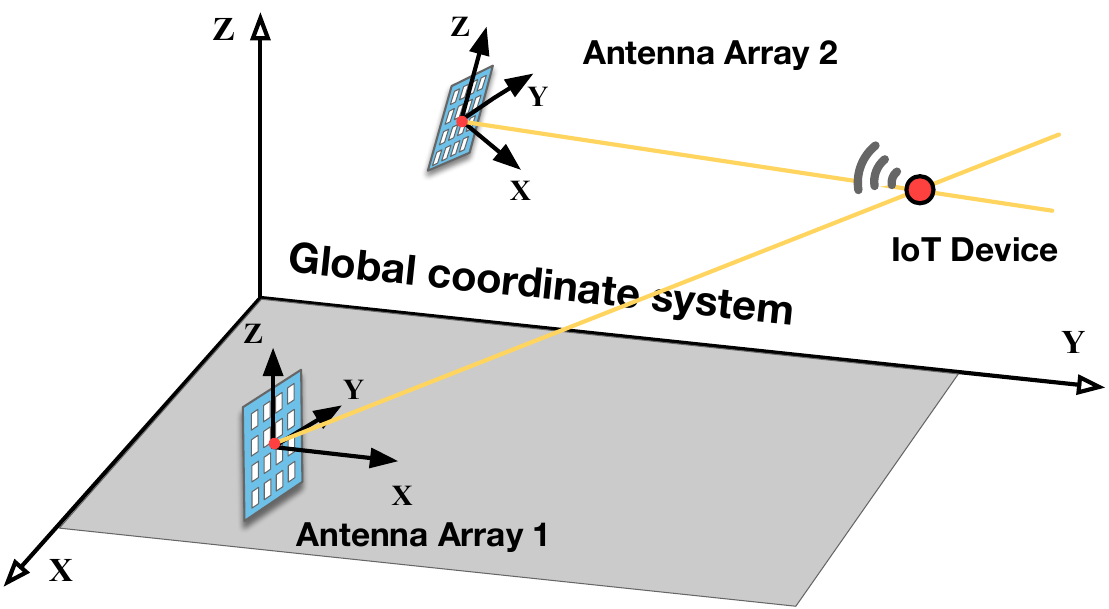}
        \label{fig:triangulation}
    }
    \caption{Triangulation-based localization estimates the position of an RF device by measuring the signal’s angle of arrival from multiple reference antenna arrays. Each array infers the direction of arrival by analyzing the spatial spectrum of the received signal.}
\end{figure}

Consider an antenna array with \( \sqrt{K} \times \sqrt{K} \) elements uniformly arranged with half a wavelength spacing between adjacent elements, as shown in Fig.~\ref{fig:ant-array}. In this configuration, we define a local Cartesian coordinate system by setting element \( E_0 \) at one corner as the origin. Using \( E_0 \) as a reference, we calculate the relative power projected in the direction \( \omega(\alpha, \beta) \) (\ie  azimuthal angle \(\alpha\) and elevation angle \(\beta\)) as follows:
\begin{equation}
\label{eqn:relative-power}
\begin{split}
\mathcal{P}(\omega) &= \frac{1}{K} \left| \sum_{k=1}^{K} w_{k}(\omega) \cdot \tilde{s}_{k} \right| \\
 &=
\begin{bmatrix}
w_{1}(\omega), w_{2}(\omega), \cdots, w_{K}(\omega)
\end{bmatrix}
\begin{bmatrix}
\tilde{s}_{1},
\tilde{s}_{2},
\cdots,
\tilde{s}_{K}
\end{bmatrix}^\top
\end{split}
\end{equation}
where \( w_{k}(\omega) = e^{\text{J} \Delta \theta_{k}} \) represents the complex weighting applied to steer a beam in the direction \( \omega \). The term \( \tilde{s}_{k} = \frac{\tilde{s}_k}{\tilde{s}_0} = e^{\text{J} \Delta \tilde{\theta}_{k}} \) signifies the ratio of the signal received by elements \( E_{k} \) and \( E_{0} \), with \( \Delta{\theta}_{k} \) and \( \Delta\tilde{\theta}_{k} \) denoting their respective theoretical and actual phase differences. The summation aggregates the relative power from all \( K \) element pairs. When the actual phase difference \( \Delta\tilde{\theta}_{k} \) matches the theoretical phase difference \( \Delta{\theta}_{k} \), the relative power \( \mathcal{P}(\omega) \) achieves its peak value. 

\begin{figure}[t!]
    \centering
    \subfigure[3D Spatial Spectrum]{
        \includegraphics[height=6cm]{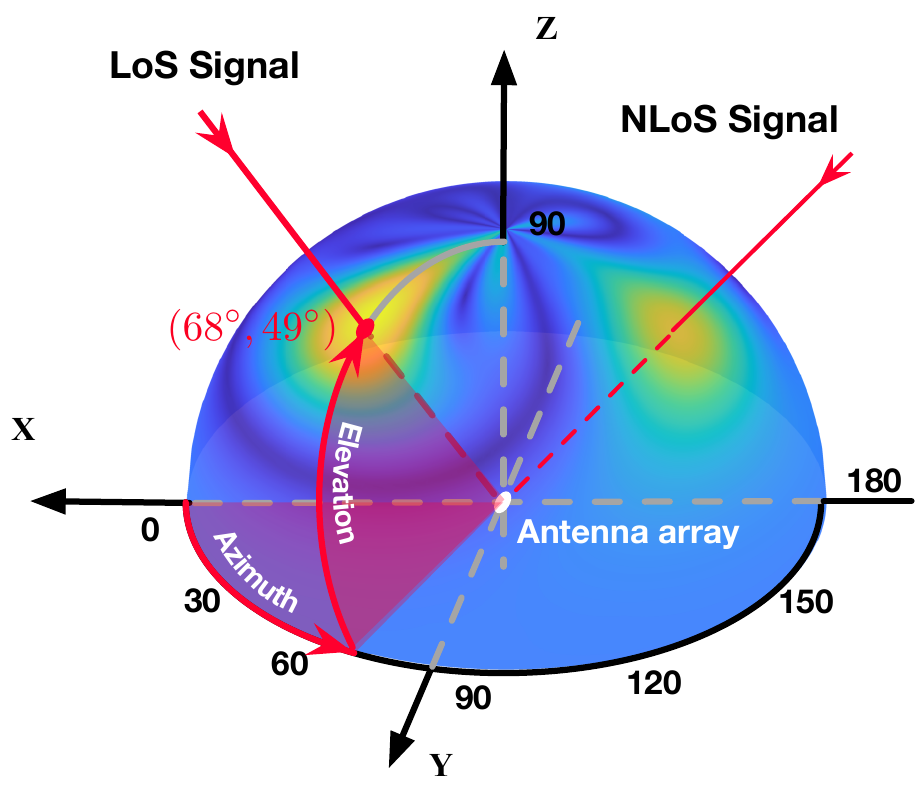}
        \label{fig:spatial-3d}
    }%
    \hspace{1.5cm}
    \subfigure[2D Spatial Spectrum]{
        \includegraphics[height=6cm]{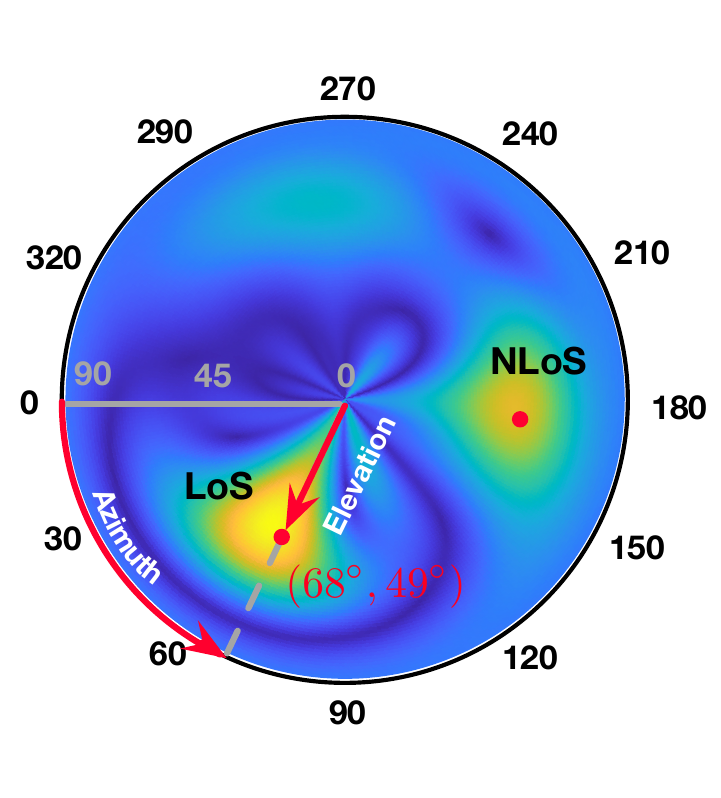}
        \label{fig:spatial-2d}
    }
    \caption{Illustration of spatial spectrum. \textnormal{On the left is the 3D spatial spectrum, while the right displays the corresponding 2D spatial spectrum.}}
    \label{fig:spatial-spectrum}
\end{figure}

\textbf{Spatial Spectrum}. Subsequently, a heatmap can be generated to depict the relative power across $N$ potential directions from which the received RF signal might originate. This representation is referred to as a \emph{spatial spectrum} and is defined as follows:
\begin{equation}\label{eqn:spatial-spectrum}
\begin{split}
	\boldsymbol{\Omega} &= \begin{bmatrix}\mathcal{P}(\omega_1) & \mathcal{P}(\omega_2) & \cdots & \mathcal{P}(\omega_N) \end{bmatrix}^T + \begin{bmatrix}Z_1 & Z_2 & \cdots & Z_N \end{bmatrix}^T\\
	&= \begin{bmatrix}
		w_{1}(\omega_1) &w_{2}(\omega_1) & \cdots &w_{K}(\omega_1)\\
		w_{1}(\omega_2) &w_{2}(\omega_2) & \cdots &w_{K}(\omega_2)\\
		\vdots & \vdots & \vdots\\
		w_{1}(\omega_N) &w_{2}(\omega_N) & \cdots &w_{K}(\omega_N)\\
	\end{bmatrix}
    \begin{bmatrix}
    	\tilde{s}_{1} \\
    	\tilde{s}_{2}\\
    	\vdots\\
    	\tilde{s}_{K}
    \end{bmatrix} + \begin{bmatrix}Z_1 \\ Z_2 \\ \cdots \\ Z_N \end{bmatrix}\\
	&=\mathbf{W}\mathbf{S} + \mathbf{Z}\raisetag{0.3cm}
\end{split}
\end{equation}
where $\mathbf{W}$, $\mathbf{S}$, and $\mathbf{Z}$ represent the weighting matrix (aka the steering matrix), the received signals, and the noise, respectively. The value of $N$ can be adjusted based on the desired angle resolution; for instance, $N=36\times 9$ would be suitable for a resolution of ten degrees. Fig.~\ref{fig:spatial-spectrum} illustrates a sample of the spatial spectrum. Specifically, Fig.~\ref{fig:spatial-3d} demonstrates the spatial spectrum in 3D, where the X and Y axes correspond to the two angles, respectively, while Fig.~\ref{fig:spatial-2d} displays a 2D projection of the 3D spectrum onto the X-Y plane. 

The spatial spectrum is expected to peak in the direction from which the RF signal originates. However, due to multipath effects, the RF signal may arrive from multiple directions, potentially resulting in multiple peaks and leading to ambiguities in pinpointing the true source. {\color{black}To address this, the true direction of arrival can be determined by finding the angle $\omega$ from $\Omega$:}
{\color{black}
\begin{equation}\label{eqn:omega*}
\omega^* = \omega_{\hat{i}}, \quad \text{where } \hat{i} = \underset{i \in \{1, \dots, N\}}{\arg\max} \ \Omega_i
\end{equation}}
This equation seeks to identify the direction that yields the most accurate real-world measurements.

\textbf{Triangulation}. Each array can pinpoint a specific direction. The position of the device is then determined either at the intersection of two such directions or at the centroid of the region formed by the overlap of multiple directions. This method of locating a position is known as \emph{triangulation}, as depicted in Fig.~\ref{fig:triangulation}. The process can be described mathematically by the following optimization problem:
\begin{equation}\label{eqn:p*}
		p^{*} = \argmin_{p} \sum_{g=1}^G \mathbf{d}(p, l(\mathbf{R}_g\cdot \vec{\omega}_g^*+\mathbf{O}_g)) 
\end{equation} 
where $\mathbf{d}(\cdot)$ represents the Euclidean distance between a position \( p \) and a line \( l \), with \( G \) being the number of antenna arrays. The term \( l(\mathbf{R}_i\cdot \vec{\omega}_i^*+ \mathbf{O}_i) \) denotes the direction (i.e., a ray) estimated by the \( i^\text{th} \) antenna array, where $\mathbf{R}_i$ is the transformation matrix converting the direction from local to global coordinate systems, and $\mathbf{O}_i$ is the array's position.

\textbf{Summary}. The overarching objective of indoor localization is to solve two complex nonlinear optimization problems outlined in Eqs.~\ref {eqn:omega*} and~\ref{eqn:p*}. This paradigm aligns naturally with the strengths of deep neural networks, which excel at addressing the optimization problem like ours. Given the proven efficacy of existing efforts in developing localization models (LMs), as evidenced by prior research~\cite{ayyalasomayajula2020deep, an2020general,li2021deep,qian2021supervised,zhan2021deepmtl}, we believe that LMs hold considerable promise for enhancing indoor localization performance.


\section{OVERVIEW}
\label{section:design}

In this section, we provide a high-level overview of the system.
\subsection{Scope}
Antenna arrays, essential to advanced communication technologies like MIMO and beamforming, are increasingly integrated into indoor localization standards, as exemplified by Wi-Fi 802.11az and Bluetooth 5.1+. These arrays enhance both signal strength and localization accuracy by enabling spatial diversity and directional signal transmission. Following this principle, our system uses antenna-array-based localization, leveraging base stations as known anchors.

\subsection{Autoencoder Architecture}

The proposed RFGP pretraining framework comprises two core components: LM and RF-NeRF, with the primary objective of pretraining the LM using unlabeled RF signals.
\begin{itemize}[leftmargin=*]
\setlength{\parskip}{0pt}
\setlength{\itemsep}{0pt plus 1pt}
\item \textbf{LM as the RFGP Encoder}. We introduce a new localization model, \oursystem, which adopts a Transformer encoder-only architecture. To effectively capture scene diversity and complexity, we integrate a Mixture of Experts (MoE) architecture into the model. Further architectural details are provided in $\S$\ref{section:lm}.
\item \textbf{RF-NeRF as the RFGP Decoder}. \textcolor{black}{The RF-NeRF component parameterizes scene-specific RF response using a radiance-field formulation. We adopt volumetric scene representations combined with voxel-based radiosity, and employ a ray-marching technique to simulate RF signal propagation from a given transmitter location. Additional details can be found in $\S$\ref{section:nerf}.}
\end{itemize}
\noindent Importantly, RFGP is a flexible pretraining framework that can be coupled with any large localization model or RF-NeRF variant. In this work, we use \oursystem and NeRF$^2$ as representative examples to demonstrate their effectiveness.

 \section{RFGP ENCODER}
\label{section:lm}

This section introduces an extended and optimized Transformer-based LM, called \oursystem.

\begin{figure}
   \centering
	\includegraphics[width=0.7\linewidth]{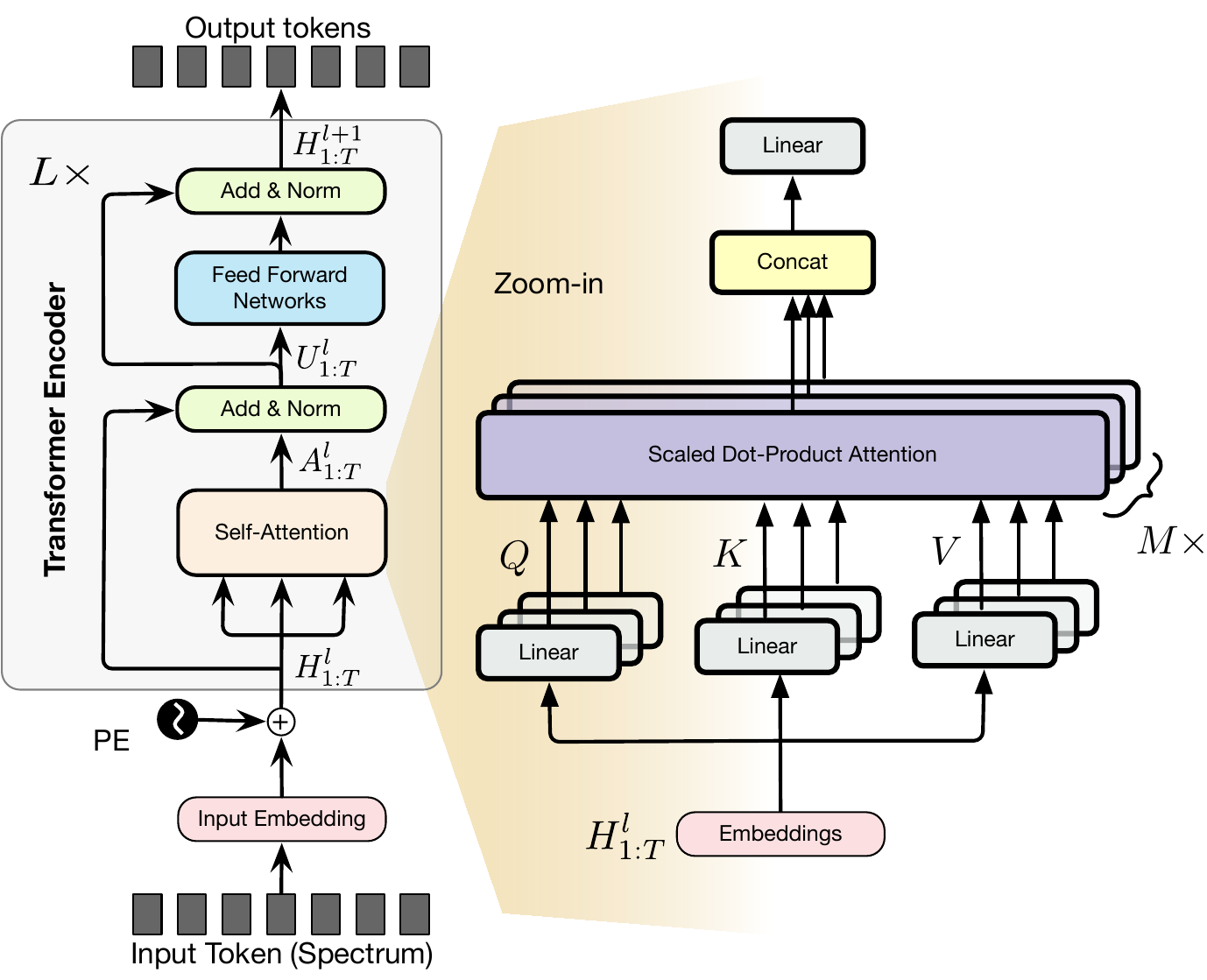}
	\caption{Illustration of Transformer Block. \textnormal {It consists of several key components, including the tokenization, positional encoding, self-attention, multi-head attention, layer normalization, and FFN.}}
	\label{fig:transformer-encoder}
\end{figure}
{\color{black}
\subsection{Transformer for Localization}
\label{subsec:transformer-localization}

Transformers have demonstrated strong scalability and representational capabilities in both language modeling and wireless localization~\cite{vaswani2017attention,zhao2024understanding}. Following this line of work, \oursystem adopts a Transformer encoder-only architecture (Fig.~\ref{fig:transformer-encoder}) to extract spatial features from RF spatial spectra. Compared with a full encoder-decoder Transformer, the encoder-only design is sufficient for our localization setting, where the goal is to map received spectra from multiple antenna arrays into a compact location-relevant representation.

Given the spatial spectrum $\Omega_i$ received by the $i$-th antenna array, where $i=1,\dots,G$, we partition each spectrum with resolution $36\times 9$ into $36$ non-overlapping $3\times3$ patches. Each patch corresponds to a $30^\circ \times 30^\circ$ directional cone. Let $\omega_{i,j}\in\mathbb{R}^{3\times3}$ denote the $j$-th patch from the $i$-th antenna array, where $j=1,\dots,36$. The patch is flattened and projected into a $d$-dimensional token embedding:
\begin{equation}
    x_{i,j}=W_{\mathrm{tok}}\operatorname{vec}(\omega_{i,j})+b_{\mathrm{tok}},
\end{equation}
where $x_{i,j}\in\mathbb{R}^{d}$ and the value of $d$ is specified in $\S$\ref{model_config}.

To encode both patch order and antenna-array position, we add a sinusoidal positional encoding and a learnable projection of the antenna-array coordinate. Specifically, let $O_i=(x_i,y_i,z_i)$ denote the 3D position of the $i$-th antenna array, and let $t=36(i-1)+j$ be the global token index. The initial token representation is
\begin{equation}
    h_{i,j}^{0}=x_{i,j}+\mathrm{PE}(t)+W_{\mathrm{pos}}O_i,
\end{equation}
where $W_{\mathrm{pos}}\in\mathbb{R}^{d\times3}$. The resulting input sequence is
\begin{equation}
    H^{0}=[h_1^0,h_2^0,\dots,h_T^0]^\top \in \mathbb{R}^{T\times d},
\end{equation}
where $T=36G$ is the total number of tokens. For example, $T=72$ for two antenna arrays and $T=108$ for three antenna arrays.

Each Transformer layer updates the token sequence through multi-head self-attention and a feed-forward network. For the $l$-th layer, where $l=1,\dots,L$, the attention output of the $m$-th head is
\begin{equation}
    A^{l,m}=\operatorname{softmax}\left(
    \frac{Q^{l,m}(K^{l,m})^\top}{\sqrt{d_h}}
    \right)V^{l,m},
\end{equation}
where
\begin{equation}
    Q^{l,m}=H^{l-1}W_Q^m,\quad
    K^{l,m}=H^{l-1}W_K^m,\quad
    V^{l,m}=H^{l-1}W_V^m.
\end{equation}
Here, $M$ is the number of attention heads and $d_h=d/M$ is the hidden dimension of each head. The outputs of all heads are concatenated and projected back to $\mathbb{R}^{d}$:
\begin{equation}
    \operatorname{MHA}(H^{l-1})
    =
    \operatorname{concat}(A^{l,1},\dots,A^{l,M})W_{\mathrm{out}}.
\end{equation}
With residual connection and layer normalization, the attention block output is
\begin{equation}
    U^{l}=\operatorname{LayerNorm}\left(
    H^{l-1}+\operatorname{MHA}(H^{l-1})
    \right).
\end{equation}
The standard Transformer layer then applies a position-wise feed-forward network:
\begin{equation}
    H^{l}=\operatorname{LayerNorm}\left(
    U^{l}+\operatorname{FFN}(U^{l})
    \right).
\end{equation}

After the final layer, we mean-pool the output tokens and feed them into a two-layer MLP to obtain the latent location feature:
\begin{equation}
    f_p=\operatorname{MLP}_{\mathrm{feat}}\left(
    \frac{1}{T}\sum_{t=1}^{T}H_t^{L}
    \right),
\end{equation}
where $f_p\in\mathbb{R}^{d_{\mathrm{feature}}}$. During fine-tuning, $f_p$ is further mapped to the predicted transmitter position by a regression head.

\subsection{Mixture of Experts}
\label{subsec:moe}

To improve model capacity while keeping the computation manageable, we integrate a Mixture-of-Experts (MoE) module into \oursystem. MoE (Fig.~\ref{fig:moe}) has been widely used in large-scale Transformer models to increase the number of parameters without activating all parameters for every input~\cite{shazeer2017outrageously,dai2024deepseekmoe,dou2023loramoe}. In our design, the MoE layer replaces the standard FFN in selected Transformer layers.

\begin{figure}
    \centering
    \includegraphics[width=0.6\linewidth]{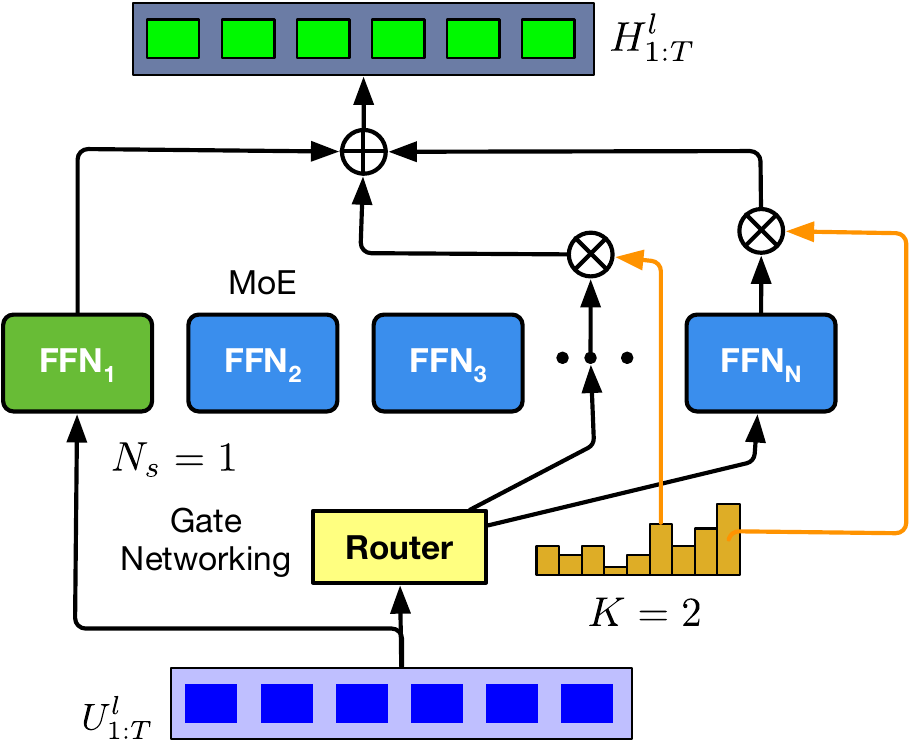}
    \caption{Illustration of Mixture-of-Experts. \textnormal{The MoE layer replaces a single FFN with $N$ expert networks $\{\operatorname{FFN}_1,\dots,\operatorname{FFN}_N\}$. The output combines $N_s$ shared experts, which process all tokens, and $K$ optional experts selected by a gating network.}}
    \label{fig:moe}
\end{figure}

Let $N$ denote the total number of experts. Among them, the first $N_s$ experts are shared experts that process every token, while the remaining $N-N_s$ experts are optional experts selected dynamically. For token $t$ in layer $l$, the MoE output is
\begin{equation}
    \operatorname{MoE}(U_t^l)
    =
    \sum_{i=1}^{N_s}\operatorname{FFN}_i(U_t^l)
    +
    \sum_{i=N_s+1}^{N}g_{i,t}\operatorname{FFN}_i(U_t^l),
\end{equation}
where $g_{i,t}$ is the gating weight of the $i$-th optional expert for
token $t$. The layer output becomes
\begin{equation}
    H_t^l
    =
    \operatorname{LayerNorm}\left(
    U_t^l+\operatorname{MoE}(U_t^l)
    \right).
\end{equation}

The gating network computes token--expert affinity scores and dynamically selects the top-$K$ optional experts for each token. Let $E_i^l\in\mathbb{R}^{d}$ be the learnable centroid embedding of the $i$-th expert in layer $l$. The unnormalized affinity score is
\begin{equation}
    a_{i,t}=(U_t^l)^\top E_i^l.
\end{equation}
\textcolor{black}{The affinity scores are normalized over all optional experts:
\begin{equation}
    s_{i,t}
    =
    \frac{\exp(a_{i,t})}
    {\sum_{j=N_s+1}^{N}\exp(a_{j,t})},
    \qquad
    i\in\{N_s+1,\dots,N\}.
\end{equation}
The selected optional-expert set is
\begin{equation}
    \mathcal{K}_t
    =
    \operatorname{TopK}_{i\in\{N_s+1,\dots,N\}}
    (s_{i,t},K).
\end{equation}
The probabilities of the selected experts are then renormalized to obtain the sparse gating weights:
\begin{equation}
g_{i,t}=
\begin{cases}
\displaystyle
\frac{s_{i,t}}
{\sum_{j\in\mathcal{K}_t}s_{j,t}},
& i\in\mathcal{K}_t, \\[2ex]
0, & \text{otherwise}.
\end{cases}
\end{equation}
Thus, each token activates all shared experts and only $K$ optional experts, improving model capacity while maintaining sparse computation.}

\textbf{Why MoE?} The MoE design is well-suited for RF localization because indoor environments exhibit diverse propagation patterns. We expect shared experts to capture common RF propagation patterns, such as directionality and attenuation-related cues, while optional experts can adapt to scene-dependent effects, such as cluttered layouts, reflective surfaces, and non-line-of-sight propagation. This conditional routing enables the model to reuse knowledge from similar environments, thereby improving cross-scene generalization.
}
%

\section{RFGP DECODER}
\label{section:nerf}

Although several \rfnerf estimation frameworks have been proposed, we adopt the \nerf framework from~\cite{zhao2023nerf2} as our \rfnerf implementation. In this section, we present the core concepts of this \nerf framework.

\subsection{Voxel Radiosity}

The proposed framework employs a voxel-based approach to model electromagnetic (EM) wave propagation in three-dimensional environments. The scene is discretized into uniform cubic voxels, with each volumetric unit capturing the spatial, attenuation, and radiative characteristics of its corresponding region. \textcolor{black}{This discretization provides a structured framework for modeling wave propagation phenomena, including signal attenuation, phase modulation, and directional scattering effects, thereby approximating EM signal behavior in complex environments.}

\textbf{Voxel Attribute Modeling}: 
Each voxel is characterized by three fundamental properties that govern its interaction with EM waves. First, the coordinates \( P_x(x, y, z) \) define the voxel's spatial location within the scene. Second, the learned attenuation coefficient \( \delta(P_x) = \Delta a(P_x) e^{\mathbf{J}\Delta \theta(P_x)} \) describes both amplitude reduction \( \Delta a(P_x) \) and phase shift \( \Delta \theta(P_x) \) as signals propagate through the voxel. Third, the directional radiation pattern \( S(P_x, \omega) \) represents the voxel's behavior as a secondary emitter, where the retransmitted signal follows \( S(P_x) = a(P_x) e^{\mathbf{J}\theta(P_x)} \) with initial amplitude \( a(P_x) \) and phase \( \theta(P_x) \). Unlike isotropic radiation sources, voxels exhibit direction-dependent scattering properties, characterized by the angular vector \( \omega = (\alpha, \beta) \) that specifies the azimuth and elevation relative to the receiver's position.

\begin{figure}[!t]
	\centering
	\includegraphics[width=0.8\linewidth]{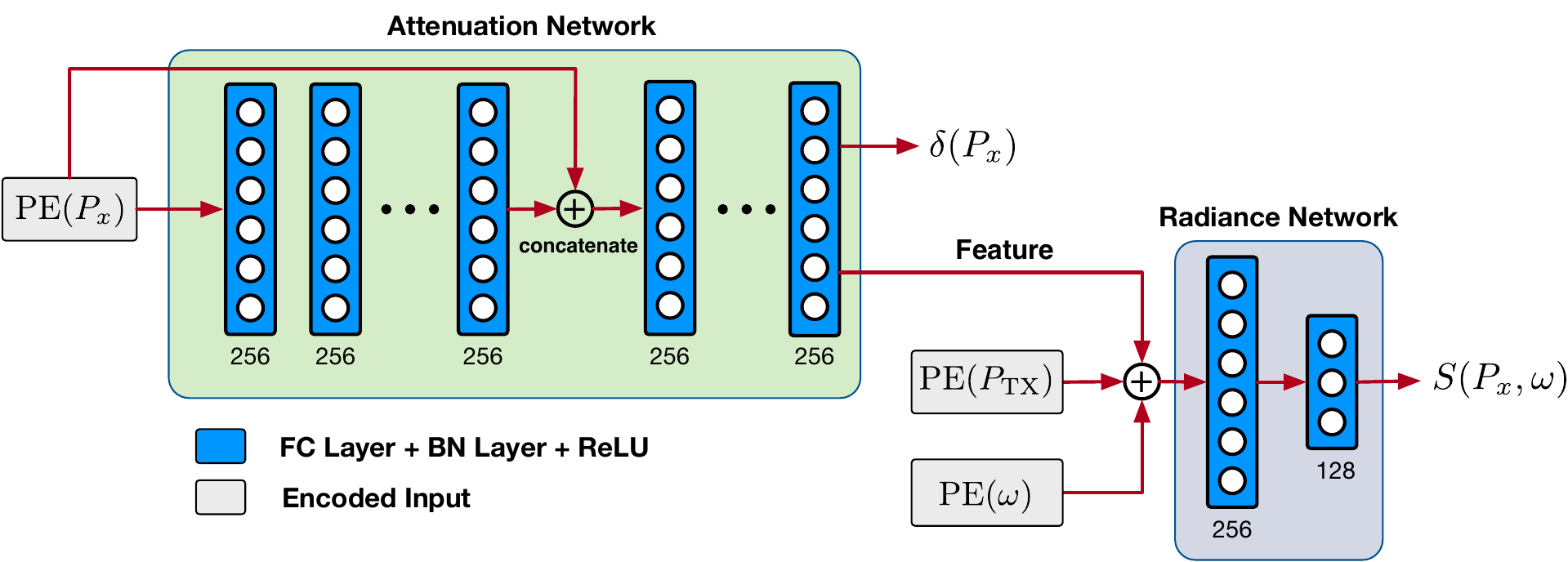}
	\caption{Architecture of \nerf . \textnormal{The network consists of two MLPs, the attenuation network, and the radiance network. The attenuation network can predict the attenuation $\delta$ of any voxel. Given the TX position and a measuring direction,  the radiance network can predict the signal transmitted from an arbitrary voxel.}}
	\label{fig:nerf}
\end{figure}

\textbf{Neural Radiance Network}:  To effectively estimate these radiation attributes, we employ a neural network \( \mathbf{F}_{\Theta} \) that constructs a continuous radiance field representation. As illustrated in Fig.~\ref{fig:nerf}, the network takes three encoded inputs: the target device's position \( P_\text{TX} \), voxel coordinates \( P_x \), and direction vector \( \omega \). Each input undergoes positional encoding \( \text{PE}(\cdot) \) to expand its dimensionality to 128, enhancing the network's ability to learn high-frequency signal variations. The network then outputs the corresponding attenuation and radiation characteristics:
\[
\mathbf{F}_{\Theta}: (P_\text{TX}, P_x, \omega) \rightarrow \left( \delta(P_x), S(P_x, \omega) \right),
\]
where \( \Theta \) denotes the trainable parameters of the neural network. This parameterization enables efficient learning of the complex relationship between spatial configuration and EM wave propagation characteristics.

\subsection{Ray Tracing}

\begin{figure}[!t]
	\centering
	\includegraphics[width=0.7\linewidth]{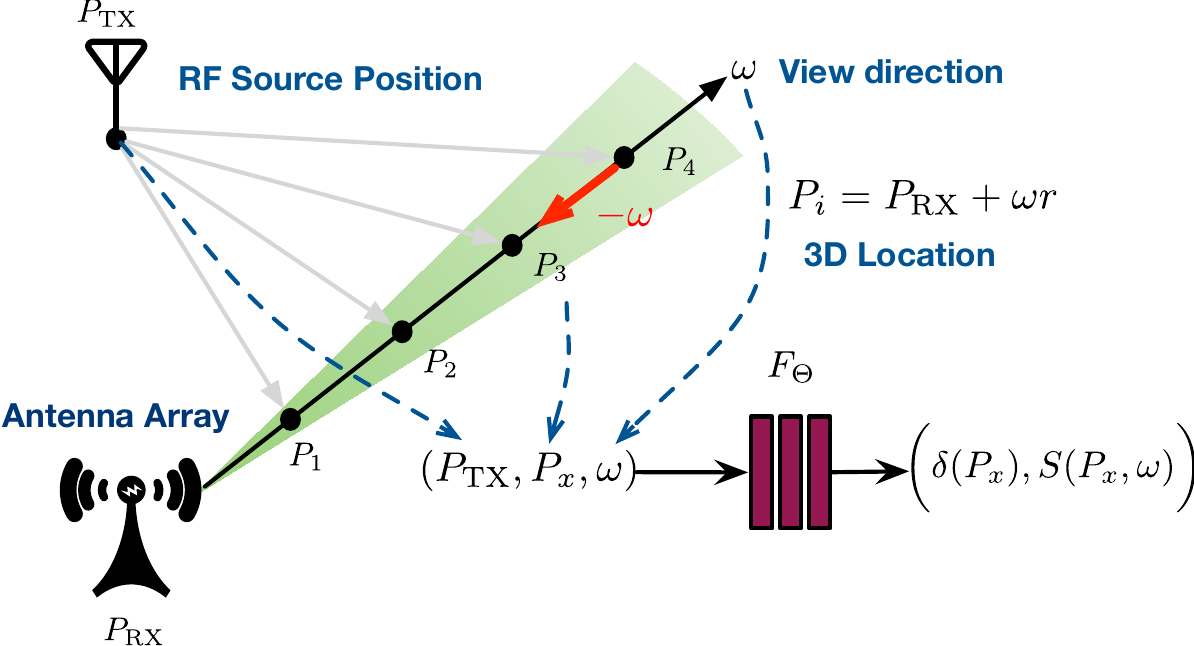}
	\caption{Illustration of ray tracing.  \textnormal{There are four voxels at $P_1 - P_4$ on the ray. Each voxel becomes a new transmitter that emits the signal along the ray to the RX. Their signals are attenuated by the other voxels between the new transmitters and the RX. }}
	\label{fig:ray-tracing}
\end{figure}

The ray-tracing process isolates and analyzes signals arriving from a specific direction $\omega$. To facilitate this analysis, we define a ray originating at the base station equipped with an antenna array (RX) and extending in direction $\omega$, as illustrated in Fig.~\ref{fig:ray-tracing}. Points along this ray are parameterized by:
\begin{equation}
	P(r, \omega) = P_{\text{RX}} + r\omega,
\end{equation}
where $r$ represents the radial distance from the base station. The total received signal $R(\omega)$ at the RX from direction $\omega$ is computed by integrating contributions from all voxels along the ray path:
\begin{equation}
	R(\omega) = \int_0^D H_{P(r,\omega)\to P_{\text{RX}}} \, S\left(P(r,\omega), -\omega\right) \, dr.
\end{equation}
where $S(P(r,\omega), -\omega)$ denotes the signal emitted by the voxel at $P(r,\omega)$ toward the RX (opposite to $\omega$), and $H_{P(r,\omega)\to P_{\text{RX}}}$ models the propagation and attenuation between the voxel and the RX. The integral accumulates the contributions along the ray up to maximum range $D$, accounting for attenuation and scattering effects introduced by the intervening medium.

The attenuation factor \(H_{P(r,\omega)\to P_{\text{RX}}}\) captures signal loss accumulated from all voxels between the RX and \(P(r,\omega)\):  
\[
H_{P(r,\omega)\to P_{\text{RX}}} = \prod_{\tilde{r}=0}^r \delta\left(P(\tilde{r}, \omega)\right),
\]  
where \(\delta\) quantifies the per-voxel attenuation. To simplify computation, we apply a logarithmic transformation, converting the product into a summation:  
\begin{equation}
\begin{split}
H_{P(r,\omega)\to P_{\text{RX}}} &= \exp\left(\int_0^r \ln\delta\left(P(\tilde{r}, \omega)\right) d\tilde{r}\right) \\
&= \exp\left(\int_0^r \hat{\delta}\left(P(\tilde{r}, \omega)\right) d\tilde{r}\right),
\end{split}
\end{equation}  
with \(\hat{\delta} = \ln\delta\). This logarithmic representation enables efficient calculation of cumulative attenuation. Substituting back into \(R(\omega)\), we obtain:  
\begin{equation}
	R(\omega) = \int_0^D \underbrace{\exp\left(\int_0^r \hat{\delta}\left(P(\tilde{r}, \omega)\right)d\tilde{r}\right)}_{\text{Attenuation Network}} \, \overbrace{S\left(P(r,\omega), -\omega\right)}^{\text{Radiance Network}} dr.
\end{equation}
Here, the attenuation network models signal decay along the propagation path, while the radiance network characterizes the emission properties of individual voxels.
\subsection{Spatial Spectrum Reconstruction}
The NeRF\(^2\) naturally predicts the power distribution of incoming signals across different directions, generating a spatial spectrum \(\Omega\) with $36\times 9$ directional zones as a matrix of signal magnitudes:  
\[
\Omega = 
\begin{bmatrix}
\parallel R(\omega_{1,1})\parallel_2 & \parallel R(\omega_{1,2})\parallel_2 & \cdots & \parallel R(\omega_{1,9}) \parallel_2 \\
\parallel R(\omega_{2,1})\parallel_2 & \parallel R(\omega_{2,2})\parallel_2 & \cdots & \parallel R(\omega_{2,9})\parallel_2 \\
\vdots & \vdots & \ddots & \vdots \\
\parallel R(\omega_{36,1})\parallel_2 & \parallel R(\omega_{36,2})\parallel_2 & \cdots & \parallel R(\omega_{36,9})\parallel_2
\end{bmatrix}.
\]  
The relative power values in \(\Omega\) are directly proportional to the true power derived from the antenna array.

\subsection{Summary}\textcolor{black}{
\nerf enables accurate prediction of signal behavior while preserving a physics-inspired structure useful for wireless signal reconstruction.} Notably, the generation of the spatial spectrum fundamentally depends on two key positions:
\begin{equation}
\Omega \to R(\omega) \to S(P(r,\omega), -\omega) \to
\begin{cases}
P_\text{TX} & \text{(unknown)}\\
P_\text{RX} & \text{(known)}
\end{cases}
\end{equation}
where $P_\text{RX}$ denotes the known position of the base station and $P_\text{TX}$ represents the unknown position of the target device. Accordingly, we can input the latent position feature into the radiance subnetwork to generate a spatial spectrum that is conditioned on the target device’s location.

\section{JOINT TRAINING}
In this section, we present the joint training methodology for the RFGP autoencoder framework.

  \begin{figure}
	\centering
	\includegraphics[width=0.8\linewidth]{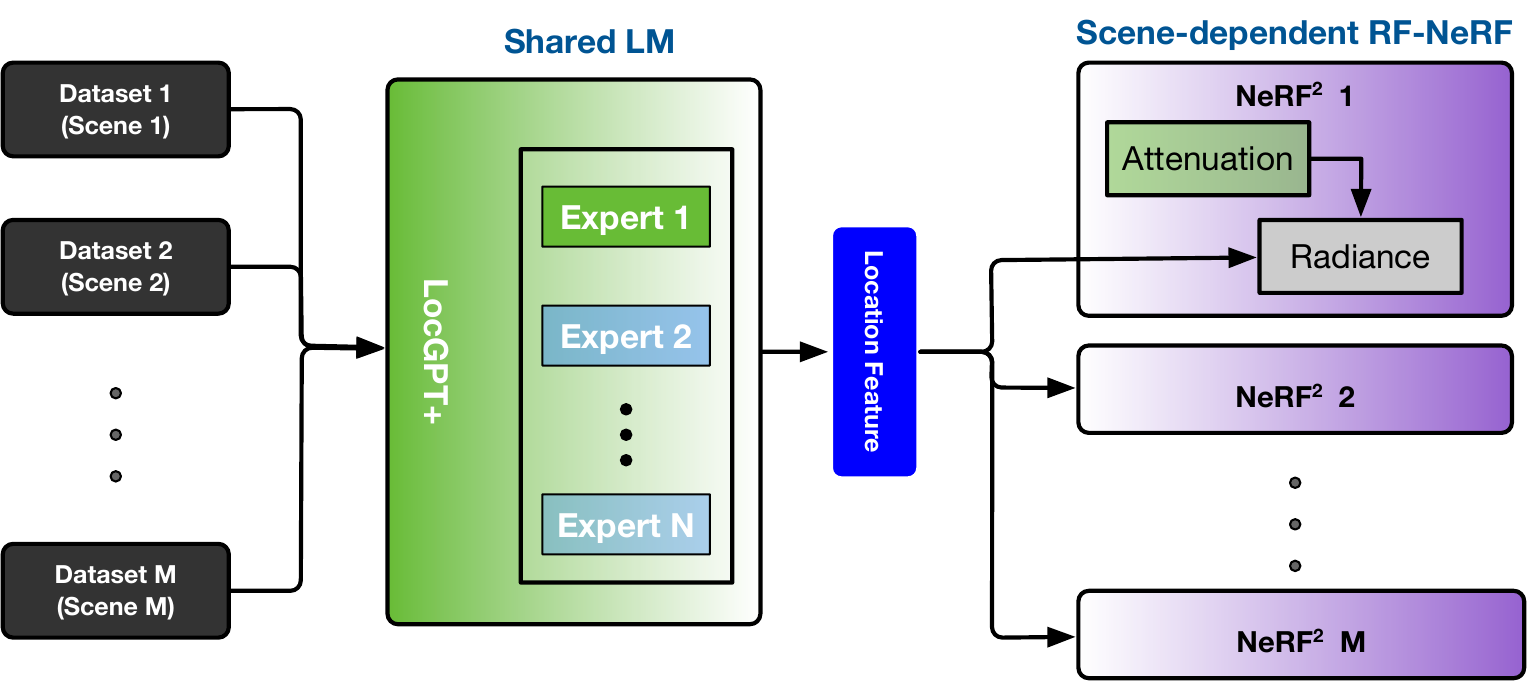}
	\caption{Illustration of the Pretraining Framework. \rfrp pretrains a scene-agnostic \llm using $M$ scene-specific \rfnerf models.}
    \label{fig:autoencoder-set}
\end{figure}

\subsection{Asymmetric Autoencoder}

Each indoor environment has unique structural and material characteristics that significantly affect radio signal propagation. Capturing these environment-specific properties thus requires modeling each scene independently. To achieve this efficiently, \rfrp introduces an \emph{asymmetric autoencoder} architecture: coupling a shared, scene-agnostic \oursystem with multiple distinct, scene-specific \nerf models, as illustrated in Fig.~\ref{fig:autoencoder-set}. \textcolor{black}{
For each scene, a dedicated \nerf instance is trained to parameterize the scene-specific RF response induced by geometry, materials, and propagation effects. }When switching to a new scene, a new \nerf instance is initialized and trained independently. In contrast,  \oursystem is shared across all scenes to extract generalizable, scene-agnostic latent features from the measured RF signals. Specifically, during training with a dataset collected from the $m^\text{th}$ scene, the shared \oursystem encodes the collected RF measurements into compact latent representations, which are then decoded by the corresponding $m^\text{th}$ \nerf instance to reconstruct the RF signals. This explicit pairing between each scene and its respective \nerf ensures accurate modeling of environment-specific details, while the shared \oursystem promotes effective feature generalization across diverse indoor environments.

\subsection{Loss Functions}

The framework is trained using three complementary objectives, each designed to address a specific aspect of the learning process:

\textbf{(1) Consistency Loss}. The consistency loss ensures input-output spectral alignment through mean squared error, preserving the fidelity of the reconstructed spatial spectra:
\begin{equation}
    \mathcal{L}_\text{cons} = \lambda_\text{cons}\sum_{i=1}^{G} \parallel \Omega_i-\widetilde{\Omega}_i \parallel_2^2
\end{equation}
where $\Omega_i$ represents the input spatial spectra (ground truth) and $\widetilde{\Omega}_i$ denotes the reconstructed spectra for the $i^{th}$ antenna array. The hyperparameter $\lambda_\text{cons}$ controls the trade-off between spectral accuracy and other objectives. This reconstruction loss is particularly critical for applications that require precise spectral matching, such as spectrum reconstruction and wireless localization.

\textbf{(2) Expert Balance Loss}. The expert balance loss prevents routing collapse in MoE architectures by promoting balanced expert utilization through two complementary terms:
{\color{black}
\begin{equation}
    \mathcal{L}_\text{bal} = \lambda_\text{bal}\sum_{i=N_s+1}^{N}\left[ \left(\frac{N-N_s}{KT}\sum_{t=1}^T \mathbb{I}_{i,t}\right)\left(\frac{1}{T}\sum_{t=1}^T s_{i,t}\right) \right]
\end{equation}}
where $T$ is the total number of tokens, $\mathbb{I}_{i,t}$ is an indicator function (1 if token $t$ selects expert $i$, otherwise 0), and $s_{i,t}$, $N$, $N_s$, and $K$ are defined as in $\S$\ref{subsec:moe}. The first component encourages a uniform distribution of expert selection frequency, while the second regularizes the magnitude of gating scores. Together, the two terms help maintain diversity in expert specialization and avoid the common pitfall of tokens disproportionately routing to only a few popular experts.

\begin{figure}
	\centering
	\includegraphics[width=0.8\linewidth]{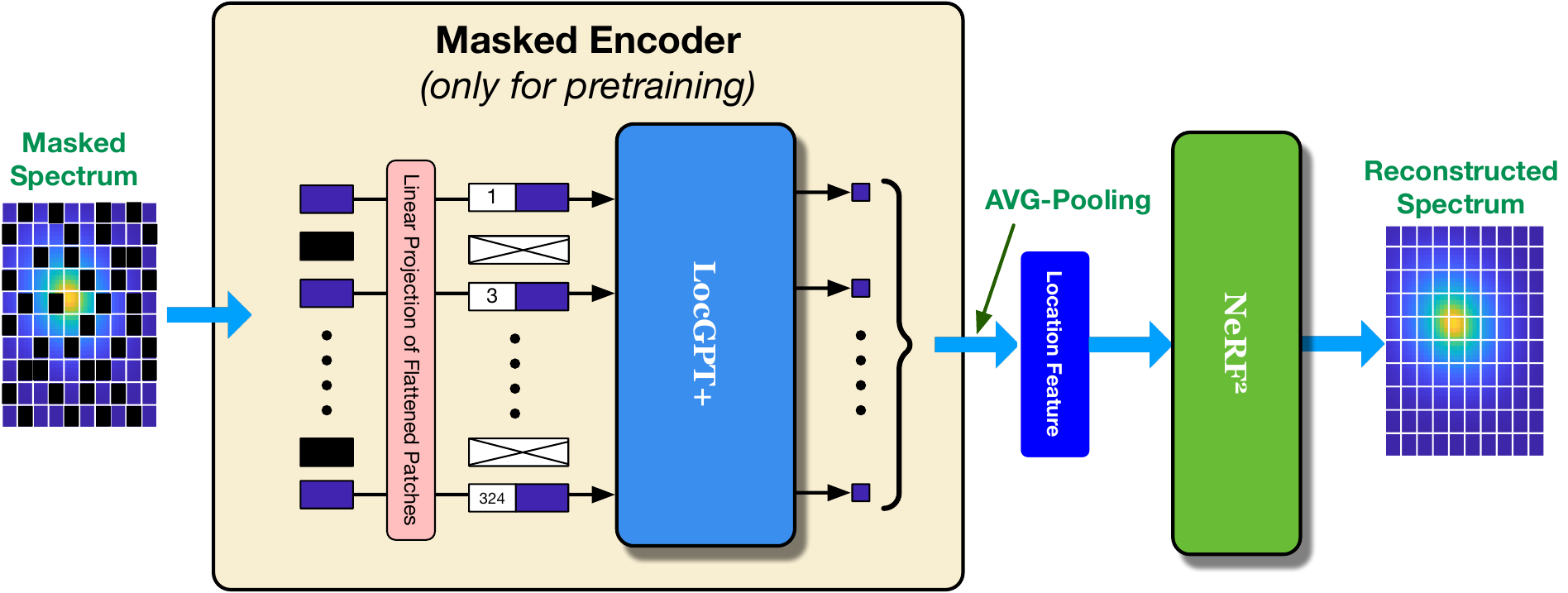}  
	\caption{Masked Encoder for Spectrum Reconstruction. \textnormal{To enhance the representation learning capability of \oursystem, we employ random patch masking on the input spectrum, forcing the model to infer missing information and generate robust latent location features.}}
	\label{fig:mae}
\end{figure}

\textbf{(3) Latent Space Regularization}. The latent space regularization encourages compact and meaningful latent representations through L2 normalization:
\begin{equation}
    \mathcal{L}_\text{lat} = \lambda_\text{lat}\parallel \mathbf{z} \parallel^2_2
\end{equation}
where $\mathbf{z}$ denotes the latent feature fed to \rfnerf. \textcolor{black}{This penalty term serves multiple purposes: it prevents arbitrary scaling of the latent space, improves numerical stability during training, and implicitly encourages compact and well-conditioned representations by favoring solutions with minimal sufficient statistics. }The regularization strength $\lambda_\text{lat}$ is typically set to a small value to avoid over-constraining the learning process.

\textbf{(4) Composite Objective}. The complete training objective combines these three loss terms through simple summation:
\begin{equation}
    \mathcal{L} = \mathcal{L}_\text{cons} + \mathcal{L}_\text{bal} + \mathcal{L}_\text{lat}
\end{equation}
During optimization, we employ gradient clipping and adaptive learning rates to handle the varying scales of these loss components. The joint optimization of reconstruction quality (via $\mathcal{L}_\text{cons}$), architectural stability (via $\mathcal{L}_\text{bal}$), and representation quality (via $\mathcal{L}_\text{lat}$) leads to robust models that generalize well across different antenna configurations and propagation environments.

\subsection{Masked Autoencoder}

RF signals are highly vulnerable to ambient interference, which often causes distortions such as hotspot displacements due to the superposition of interfered or reflected signals. To mitigate this, we adopt a Masked Autoencoder (MAE)~\cite{he2022masked}, which enhances the encoder’s robustness to noise and interference. As illustrated in Fig.~\ref{fig:mae}, a large portion of the input spectrum (e.g., 75\% of patches) is randomly masked. The input tokens corresponding to these masked patches are omitted from the encoder input (\oursystem). Because positional encodings are applied before masking, removing these tokens does not disrupt the spatial relationships among the remaining patches. The masking strategy forces the encoder to infer these correlations from incomplete data, thereby encouraging it to learn more robust and generalizable spatial-spectral representations. The latent positional features derived from the masked spectrum are then passed to the \nerf decoder, which reconstructs the full spectrum. The pretraining objective is to ensure that the reconstructed spectrum closely matches the original, unmasked spectrum. 

\begin{figure}
    \centering
    \includegraphics[width=0.8\linewidth]{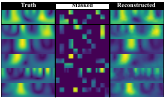}
    \caption{Illustration of Spectra Reconstruction. \textnormal{For each triplet in a row, we display the ground truth (left), the masked input spectrum (middle), and the reconstructed spectrum (right). The masking ratio is 75\%.}}
    \label{fig:reconstruction}
\end{figure}

To visually illustrate the effect of masking, Fig.~\ref{fig:reconstruction} presents six representative examples. Each row displays the original spectrum (ground truth), the masked spectrum with 75\% of patches removed, and the reconstructed spectrum produced by the autoencoder. This masking-based pretraining strategy enhances the model’s robustness to interference and promotes the learning of transferable spatial-spectral features essential for accurate downstream localization.

\subsection{Supervised Fine-Tuning}

For the downstream localization task, we fine-tune the pretrained model using a small amount of labeled data collected from the target scene. Specifically, two or three RF spectra—each captured by a different base station—are sequentially fed into \oursystem. The model extracts a location feature from each input, and these features are concatenated into a single vector (i.e., $f_p^1 \parallel f_p^2 \parallel f_p^3$). This concatenated representation is then passed through a two-layer MLP to regress the transmitter’s coordinates $\mathbf{P}$. This procedure is formalized as follows:
\begin{equation}
	\mathbf{P}= W_\text{FT}\cdot (f_p^1 \parallel f_p^2 \parallel f_p^3)
\end{equation}
The fine-tuning objective minimizes the Euclidean distance between the predicted and ground-truth positions:
\begin{equation}
\mathcal{L}_\text{fine-tune} = \parallel \mathbf{P} - \mathbf{P}^* \parallel_2^2,
\end{equation}
where $\mathbf{P} \in \mathbb{R}^3$ denotes the predicted transmitter coordinates and $\mathbf{P}^*$ represents the ground-truth location.

{\color{black}
\section{IMPLEMENTATION}
\label{sec:implementation}
\begin{table}[t!]
\centering
\caption{Summary of Training Datasets}
\label{tab:datasets}
\setlength{\tabcolsep}{0pt} 
\begin{threeparttable}
\begin{tabular*}{\textwidth}{@{\extracolsep{\fill}} c l c c c c c @{}}
\toprule
\textbf{\#} & \textbf{Tech.} & \textbf{Freq. (GHz)} & \textbf{Scene (\#)} & \textbf{Samples (\#)} & \textbf{Station (\#)} & \textbf{Density} \\
\midrule
1 & RFID & 0.920      & 28 & 1,303,710 & 3 & 7,227 \\
2 & BLE  & 2.4        & 29 & 4,344,171 & 3 & 3,027 \\
3 & IIoT & 1.27/3.44  & 19 & 445,817   & 4 & 119   \\
4 & WiFi & 2.4        & 24 & 1,233,623 & 4 & 4,104 \\
\bottomrule
\end{tabular*}
\begin{tablenotes}
    \item[*] \footnotesize The density is represented in the unit of samples per cubic meter.
\end{tablenotes}
\end{threeparttable}
\end{table}
This section details the implementation of \oursystem, encompassing the dataset composition, data collection apparatus, model architecture, and pretraining configurations.

\subsection{Dataset and Collection Apparatus}
To ensure the model learns a robust, generalizable physical prior, we constructed a massive-scale dataset spanning diverse environments and wireless modalities. 

\textbf{Target Scenarios and Statistics.} We collected 7,327,321 RF signal samples across 100 diverse scenes, including office rooms, classrooms, restaurants, industrial warehouses, corridors, shopping malls, halls (conference halls, auditoriums), libraries, laboratories, server rooms, hospitals (patient rooms, waiting areas), stairwells, and more. Detailed statistics are provided in Table~\ref{tab:datasets}. The dataset spans multiple technologies: RFID (920 MHz), WiFi (2.4 GHz), BLE (2.4 GHz), and industrial IoT (IIoT) (1.27 GHz and 3.44 GHz). We utilize 6,687,272 samples from 75 scenes (P1-P75) for self-supervised pretraining and 640,049 samples from the remaining 25 scenes (S1-S25) for downstream evaluation. Notably, while 21.3\% of the pretraining samples have positional labels, \oursystem operates entirely unsupervised during pretraining, using no label information.

\textbf{RFID Deployment.} The deployment of our RFID localization platform is depicted in Fig.~\ref{fig:rfid}. Our design incorporates a dual-channel station with a $4\times4$ antenna array. We use a USRP X310 SDR from NI~\cite{usrp_x310} for baseband signal processing. Each X310 is equipped with two TwinRX daughterboards: one RX channel is linked to the antenna array, while the other connects to a circularly polarized patch antenna serving as the side channel. The RF source is fixed approximately 5 m from the tag, powering it and prompting the continuous transmission of \texttt{RN16} packets. A key advantage of this bistatic design is that the three antenna arrays can be placed at a considerable distance from the tag (e.g., up to 50 m). Using this platform, we successfully collected spatial data across 28 distinct scenes. The location labels are obtained using a high-precision infrared positioning system, OptiTrack~\cite{optitrack_primex13}.

\textbf{BLE Deployment.} The BLE localization platform, illustrated in Fig.~\ref{fig:ble}, was developed using Silicon Labs components operating at 2.4 GHz in compliance with BLE V4.2 specifications. The system features three $4\times4$ patch antenna arrays (model: BRD4191A)~\cite{siliconlabs2021}, each measuring $16\times16 \text{ cm}^2$ with individual patch antennas sized at $2.42\times 2.42 \text{ cm}^2$. These arrays achieve average angular errors of $\pm 3.2^\circ$ in azimuth and $\pm 4^\circ$ in elevation. Each antenna features dual-input ports for receiving both horizontally and vertically polarized signals. The array boards, constructed from JL-2800 laminate, were rigorously optimized and tested using IT-180A and IS400 laminates to enhance performance. Phase measurements were extracted via the Gecko SDK 4.1 and RTL library. The active Bluetooth tags (model RD4184A) were also sourced from Silicon Labs. The location labels are obtained using Velodyne VLP-16 LiDAR.

\textbf{IIoT Deployment.} We integrate the DICHASUS dataset released by \cite{dichasus2021} into the dataset. The DICHASUS dataset collection contains CSI datasets for line-of-sight, non-line-of-sight, indoor, outdoor, colocated, and distributed antenna configurations, complete with ground-truth positioning labels acquired by a 3D Tachymeter.

\textbf{WiFi Deployment.} The WiFi dataset consists of three components: DLoc~\cite{ayyalasomayajula2020deep}, ESPARGOS~\cite{dataset-espargos-0001,dataset-espargos-0002,dataset-espargos-0005,dataset-espargos-0007}, and our own contribution. DLoc focuses on positioning a WiFi receiver using Channel State Information (CSI), with each station featuring a four-antenna linear array. The ESPARGOS datasets are derived from extensive experience in providing datasets alongside DICHASUS, based on ESP32. 
Our system utilizes a 2×4 rectangular antenna array, built with two Ettus Research USRP X310 units~\cite{usrp_x310}. Each unit is equipped with four antennas and operates with a synchronized external clock. The system operates at 2.4 GHz with a 20 MHz bandwidth. The antenna elements are spaced 6.25 cm apart horizontally and vertically, ensuring the spacing is less than half the wavelength to prevent spatial aliasing. 
We achieved centimeter-level positioning accuracy by using a synchronized CHCNAV S10 Pocket-Sized IMU-RTK GNSS~\cite{chcnav_s10} mounted at the centroid of the array, which provides position updates at a frequency of 1 Hz.

\begin{figure*}[t]
    \centering
        \begin{minipage}[t]{0.583\linewidth}
        \centering
        \includegraphics[width=\linewidth]{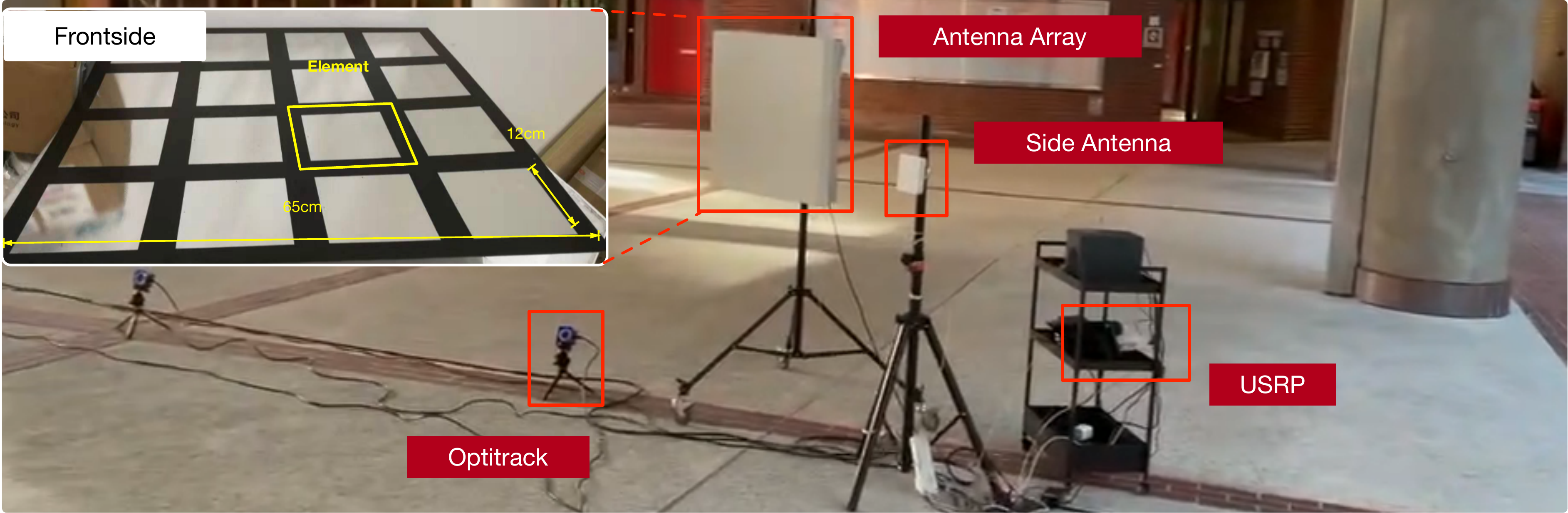}
        \caption{{\color{black}RFID Localization Deployment.}}
        \label{fig:rfid}
    \end{minipage}%
    \begin{minipage}[t]{0.417\linewidth}
        \centering
        \includegraphics[width=\linewidth]{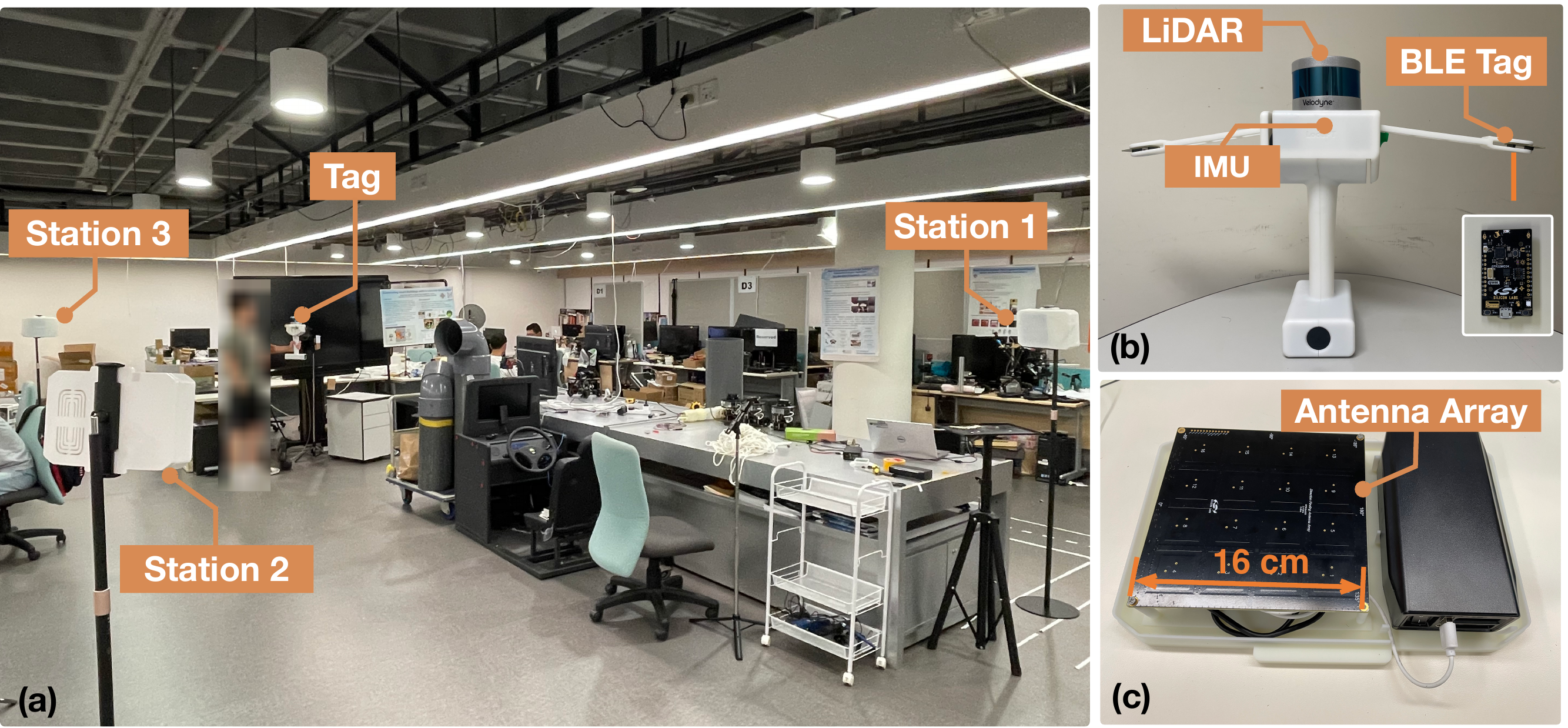}
        \caption{{\color{black}BLE Localization Deployment}}
        \label{fig:ble}
    \end{minipage}
\end{figure*}

\subsection{Model Configuration}
\label{model_config}
The \oursystem framework consists of a large localization encoder and a physics-informed spatial decoder.

\textbf{LM Configuration.} The base Large Localization Model (LM) is instantiated as a 12-layer Transformer encoder comprising 570 million parameters. We set the embedding dimension $d=1024$, the number of multi-heads to 8, and the MLP expansion ratio to 4.0. To efficiently enhance representational capacity, the 4th, 8th, and 12th layers use a Mixture-of-Experts (MoE) architecture. Each MoE layer contains 16 experts (one shared and 15 optional). A top-2 routing mechanism enables $C(15,2)=105$ potential scene-specific gating combinations. This hierarchical design progressively captures environmental complexity: mid-level spatial relationships in the 4th layer, structural patterns in the 8th layer, and high-level scene abstractions in the 12th layer. 

\textbf{\rfnerf Configuration.} We adopt \nerf~\cite{zhao2023nerf2} as the physics-guided decoder to reconstruct spatial spectra. Specifically, the attenuation subnetwork consists of eight fully connected layers (256 nodes per layer with ReLU activations), yielding a density scalar $\delta(P_x)$ and a 256-dimensional geometric feature vector. This vector is subsequently concatenated with the RX viewing direction $ \omega $ relative to $ P_x $ and the TX position $ P_\text{TX} $. This fused representation is passed into the radiance network—comprising two fully connected layers (256 and 128 nodes, ReLU activated)—which outputs the direction-dependent RF signal $ S(P_x, \omega) $ retransmitted from the voxel along direction $ \omega $. 

\subsection{Pretraining Settings and Computational Cost}
During the pretraining phase, we use the Adam optimizer with $\beta_1=0.9$, $\beta_2=0.999$, and a weight decay of 0.001. The learning rate follows a warmup-decay schedule: it increases linearly from $3 \times 10^{-5}$ to $3 \times 10^{-4}$ over the first 50 epochs, followed by a cosine decay back to $3 \times 10^{-5}$. The composite loss function is balanced using coefficients $\lambda_\text{cons} = 1$, $\lambda_\text{bal} = 0.01$, and $\lambda_\text{lat} = 0.01$. 

\textcolor{black}{
To improve training efficiency and reduce memory consumption, we use mixed-precision training during pretraining. The pretraining spans 500 epochs. Although the model has 570M parameters, the sparse routing mechanism in the MoE layers activates only a subset of experts per input, reducing the effective computational and memory footprint. }In our implementation, these optimizations enable pretraining on a single GPU server equipped with 7 NVIDIA A100 PCIe 80GB GPUs. The complete pretraining process takes approximately 150 hours and incurs an estimated computational cost of 1.2k USD.

}

\section{EVALUATION}
\label{section:evaluation}

This section evaluates the performance of \rfrp and \oursystem across a large dataset.
\begin{figure*}[t]
    \centering
    \includegraphics[width=\linewidth]{pretrain-exp1}
    \caption{{\color{black}Efficacy of Pretraining. Comparison between \oursystem fine-tuned from pretraining and a version trained from scratch, evaluated across four scenes using varying proportions of labeled training data (x-axis). The shaded regions represent the 95\% confidence intervals.}}
    \label{fig:ftrt-exp}
\end{figure*}

\subsection{Efficacy of Pretraining}

To quantify the impact of pretraining on localization performance, we compare two versions of \oursystem: one pretrained with the \rfrp framework and the other trained from scratch. Both models are trained (or fine-tuned) using varying proportions of labeled data (20\%, 40\%, 60\%, and 80\%), while evaluation is consistently conducted on the same held-out 20\% test set. The data are drawn from four test scenes—S1, S10, S20, and S25. Localization accuracy is measured by the Euclidean distance between the predicted and ground-truth transmitter positions.

The results are presented in Fig.~\ref{fig:ftrt-exp}, from which we derive three key observations:
\begin{itemize}[leftmargin=*]  
  \setlength{\parskip}{0pt}
    \setlength{\itemsep}{0pt plus 1pt}
  \item First, as expected, localization errors generally decrease as the proportion of labeled training data increases. For example, in Scene S1, the error is reduced by approximately 0.45cm and 0.55cm per additional 1\% of training data for the pretrained and non-pretrained models, respectively.
  \item Second, \oursystem with pretraining consistently outperforms its non-pretrained counterpart across all scenes. With only 20\% of the training data, pretraining yields notable reductions in localization errors: 19.6\% (S1), 17.4\% (S10), 14.1\% (S20), and 29.9\% (S25).
  \item Third, the performance gain from pretraining peaks when 60\% of the training data is used, resulting in error reductions of 23.0\% (S1), 20.0\% (S10), 47.0\% (S20), and 16.8\% (S25) compared to the non-pretrained model.
\end{itemize}

These findings validate the effectiveness of the pretraining strategy for RF localization. Pretraining enables the model to acquire generalizable spatial-spectral representations from large-scale, unlabeled wireless data. Moreover, the learned features enhance out-of-distribution generalization and mitigate overfitting, particularly in low-data regimes.

{\color{black}
\subsection{Efficacy of RFGP}
\begin{figure*}[t]
	\centering
		\begin{minipage}[t]{0.5\linewidth}
		\centering
		\includegraphics[width=\linewidth]{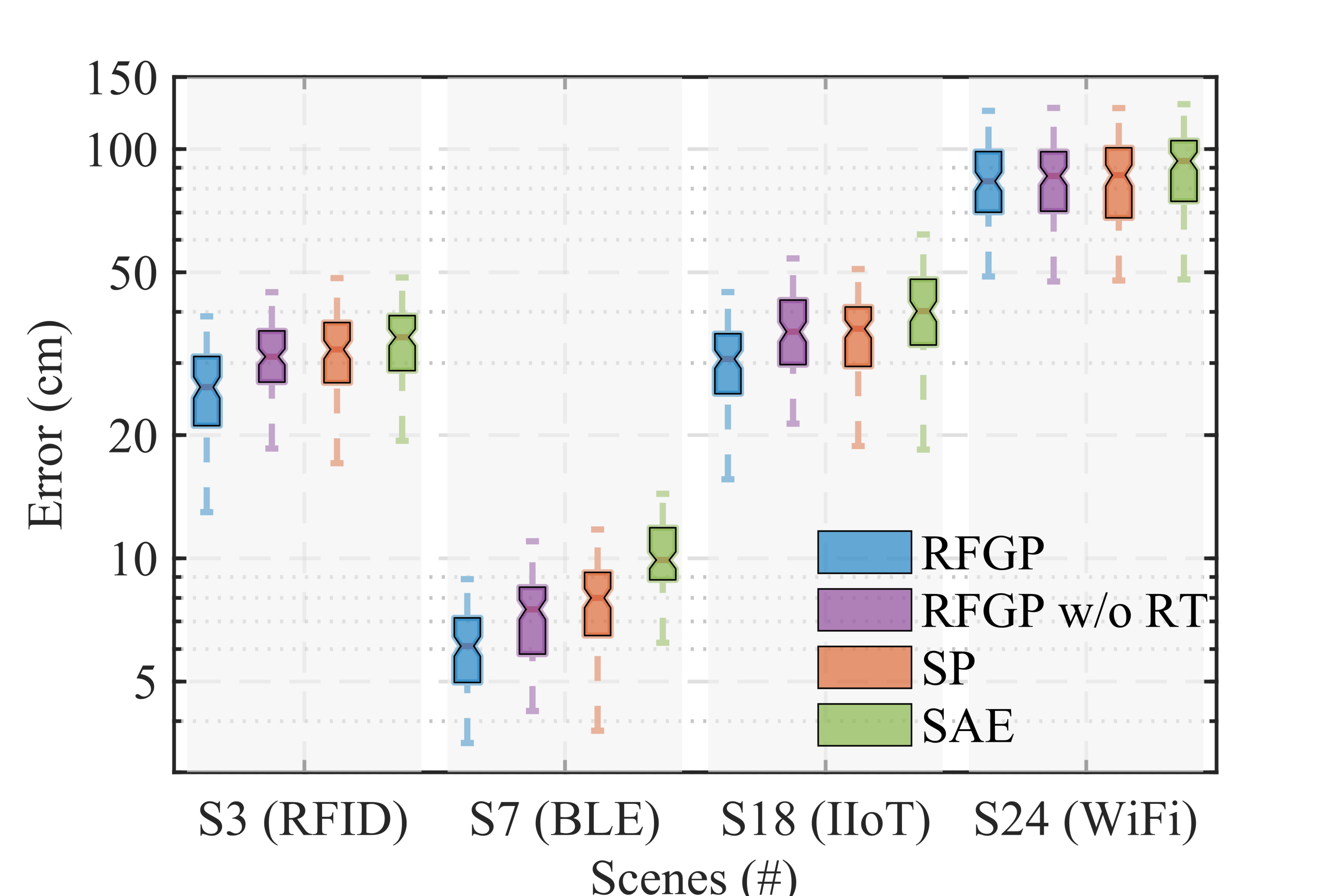}
		\caption{{\color{black}Efficacy of RFGP}}
		\label{fig:nerf2-exp}
	\end{minipage}%
	\begin{minipage}[t]{0.5\linewidth}
		\centering
		\includegraphics[width=\linewidth]{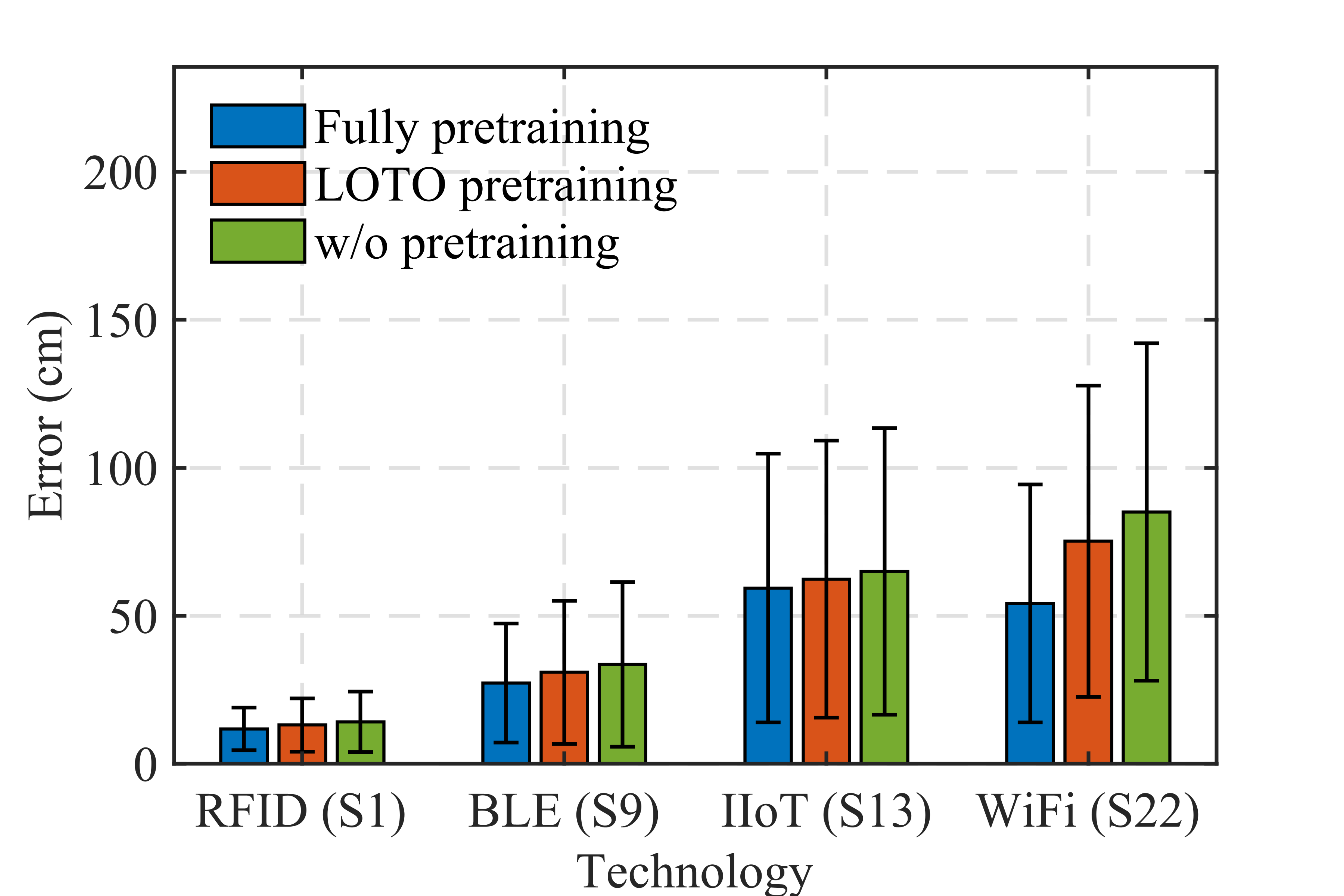}
		\caption{{\color{black}Efficacy of Cross-Technology Generalization}}
		\label{fig:loto-exp}
	\end{minipage}
\end{figure*}
Next, we evaluate the effectiveness of \rfrp by comparing it against three alternative pretraining strategies: Supervised Pretraining (SP), the Symmetrical Autoencoder (SAE), and a rigorously controlled ablation baseline, denoted as \rfrp w/o RT (RFGP without Ray-Tracing). 

Specifically, SP pretrains the \oursystem model using only 21.3\% of the pretraining set. SAE connects \oursystem to a reversed copy of itself, forming a symmetrical architecture where the shared decoder reconstructs the original spectrum. To strictly isolate the contribution of the physics-based prior, \rfrp w/o RT utilizes the exact same scene-specific decoder architecture as our full \rfrp. However, it completely bypasses the physical ray-tracing and voxel radiosity formulations, acting as a ``black box'' that directly regresses the spatial spectrum. All unsupervised methods (\rfrp, \rfrp w/o RT, and SAE) are pretrained using 100\% of the available pretraining data. After pretraining, the models are fine-tuned and evaluated on S3 (RFID), S7 (BLE), S18 (IIoT), and S24 (WiFi).

As shown in Fig.~\ref{fig:nerf2-exp}, \rfrp consistently outperforms the baseline methods across all evaluated scenarios. Specifically, \rfrp achieves the best results, recording 26.2 cm in S3 (RFID), 6.1 cm in S7 (BLE), 30.7 cm in S18 (IIoT), and 83.5 cm in S24 (WiFi). These results significantly improve upon the strongest baseline, \rfrp w/o RT, which recorded 31.1 cm, 7.5 cm, 35.8 cm, and 86.0 cm, respectively, as well as the performance of SP and SAE.

This superior performance can be attributed to three main factors. First, unsupervised pretraining enables \oursystem to leverage a much larger volume of unlabeled data, thereby facilitating the discovery of latent features. This contributes to the performance gap between unsupervised models (such as \rfrp w/o RT and \rfrp) compared to SP. Second, scene-specific decoders, as employed in both \rfrp w/o RT and \rfrp, are better at capturing the unique spectral characteristics of specific environments than a globally shared decoder, like SAE. \textcolor{black}{Finally, and most importantly, the performance gap between \rfrp and the baseline \rfrp w/o RT suggests that the ray-tracing-based reconstruction prior contributes beyond decoder capacity alone. This result supports the usefulness of physics-guided pretraining for downstream localization.}}




{\color{black}
\subsection{Efficacy of Cross-Technology Generalization}
To evaluate the model's robustness to modality shifts and its capacity for cross-technology transfer, we conduct a Leave-One-Technology-Out (LOTO) experiment. Under this protocol, data from a specific target technology is entirely withheld during the unsupervised pretraining phase and introduced only during the downstream fine-tuning stage. We benchmark LOTO pretraining against two conditions: training from scratch and fully pretraining with all technologies involved.

As detailed in Fig.~\ref{fig:loto-exp}, which reports both mean errors and standard deviations, the results across four representative scenes (S1, S9, S13, and S22) exhibit a strict performance hierarchy: Fully Pretraining $>$ LOTO Pretraining $>$ w/o Pretraining. Crucially, the LOTO strategy establishes a robust advantage over the train-from-scratch baseline. For instance, despite lacking exposure to WiFi signals during pretraining, LOTO reduces localization error on S22 by 11.6\%, reducing it from 85.1 cm to 75.2 cm. Similar generalization gains are consistently observed across BLE (8.0\% error reduction), RFID (7.7\%), and IIoT (4.0\%). 
 \textcolor{black}{
This consistent margin suggests that \rfrp learns representations with some modality-agnostic and geometry-related transferability, and that the encoder may learn spatial features that partially transfer across disparate frequency bands under the RF-NeRF decoder's ray-tracing-based constraint. While the Fully Pretraining strategy naturally achieves the highest accuracy by capturing frequency-specific attenuation characteristics, LOTO's competitive performance supports the feasibility of RFGP as a generalizable pretraining framework for heterogeneous indoor localization.}
}
\begin{figure*}
		\begin{minipage}[t]{0.33\linewidth} 
		\centering
		\includegraphics[width=\linewidth]{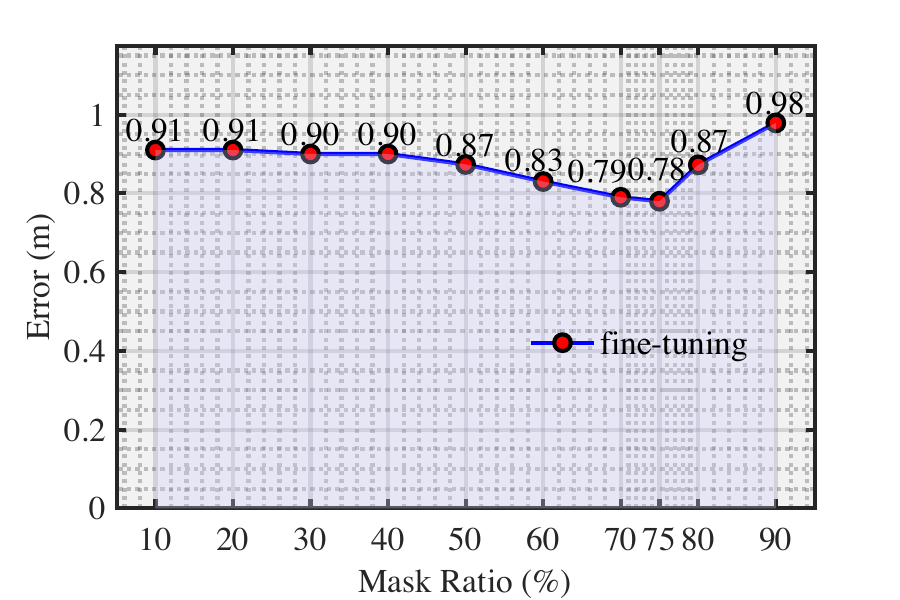}
		\caption{Optimal Masking Ratio}
		\label{fig:mask_ratio}
	\end{minipage}%
	\begin{minipage}[t]{0.33\linewidth}
		\centering
		\includegraphics[width=\linewidth]{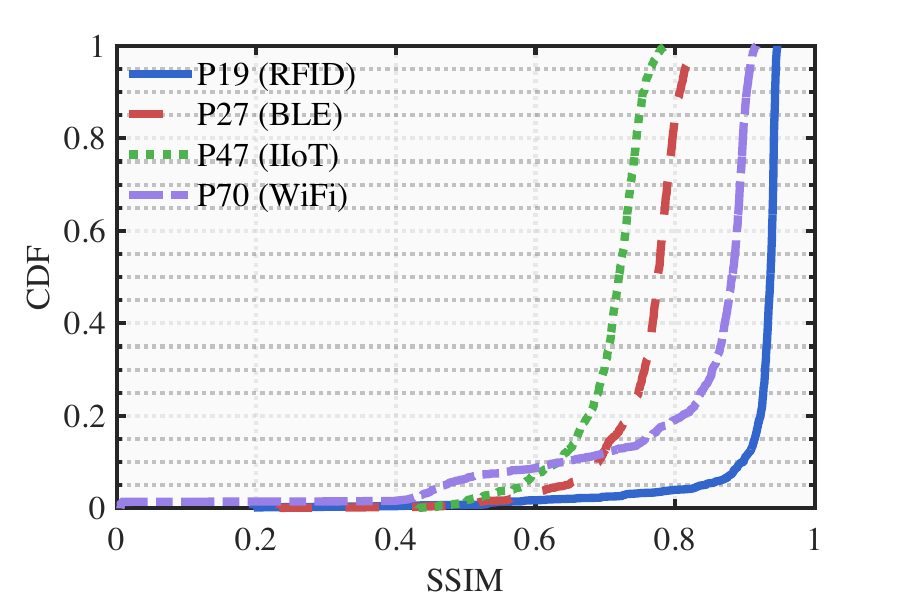}
		\caption{Efficacy of Reconstruction}
		\label{fig:ssim-exp}
	\end{minipage}%
	\begin{minipage}[t]{0.33\linewidth}
		\centering
		\includegraphics[width=\linewidth]{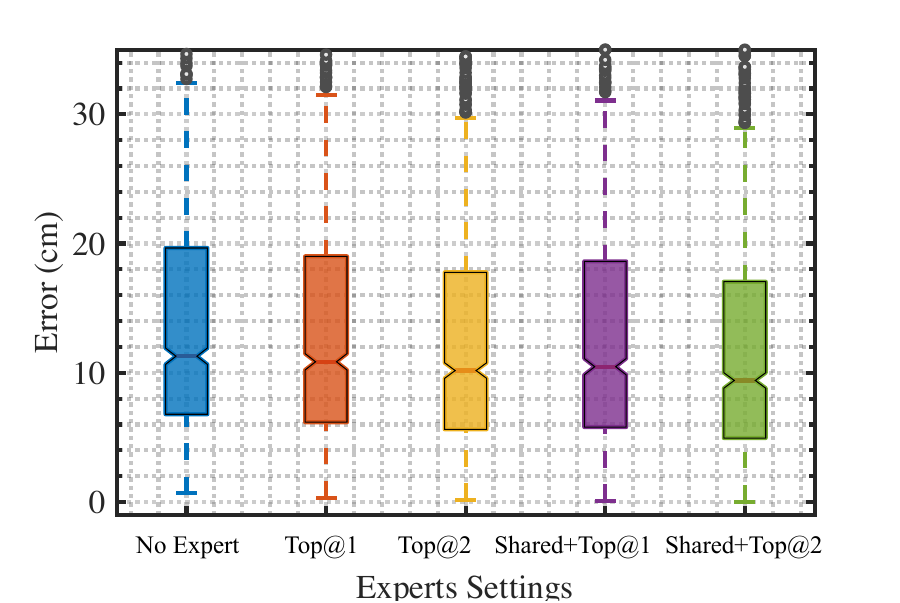}
		\caption{Efficacy of MoE}
		\label{fig:moe-exp}
	\end{minipage}
\end{figure*}
\subsection{Efficacy of Masking}


To improve the model’s robustness against noise and interference, we employ a masked autoencoder strategy in which a portion of the input spatial spectrum patches is randomly masked during pretraining. To determine the optimal masking ratio, we evaluate the localization performance of \oursystem, fine-tuned from pretrained models, on scene S12 using masking ratios ranging from 10\% to 90\%. 

As shown in Fig.~\ref{fig:mask_ratio}, localization accuracy improves with increasing masking ratios, peaking around 70\%–75\%. This improvement is attributed to the model's ability to learn stronger spatial correlations among unmasked patches, thereby enhancing generalization. However, beyond this threshold, excessive information loss impairs learning and degrades performance. Based on these results, a 75\% masking ratio is recommended for optimal accuracy.

Using this optimal ratio, we further evaluate reconstruction quality. Fig.~\ref{fig:ssim-exp} shows the cumulative distribution function (CDF) of the structural similarity index (SSIM) between the original and reconstructed spectra across four pretraining scenes. Specifically, \rfrp achieves mean SSIM values of 0.94, 0.78, 0.72, and 0.88 for pretraining scenes P19, P27, P47, and P70, respectively. These high SSIM scores demonstrate that the model effectively preserves critical spectral structure, enabling accurate feature recovery and supporting improved downstream localization performance.

\subsection{Efficacy of MoE}

Next, we conduct an ablation study to examine the impact of the MoE architecture on localization accuracy. The study is performed on the S23 scene, comparing four different MoE configurations: no expert, Top@1 expert, Top@2 experts, a shared expert plus Top@1 optional expert, and a shared expert plus Top@2 optional experts. 

The results are shown in Fig.~\ref{fig:moe-exp}. The figure yields several key insights:
\begin{itemize}[leftmargin=*]  
  \setlength{\parskip}{0pt}
    \setlength{\itemsep}{0pt plus 1pt}
  \item First, integrating the MoE architecture consistently improves accuracy, with gains ranging from $3.5\%$ to $16.8\%$, highlighting the effectiveness of MoE in enhancing performance.
  \item Second, adding a shared expert further improves accuracy regardless of whether Top@1 or Top@2 experts are selected. Specifically, the shared expert contributes approximately $3.6\%$ and $7.8\%$ median error reduction for the Top@1 and Top@2 settings, respectively.
  \item Third, the Top@2 configuration outperforms the Top@1 configuration, achieving improvements of $9.7\%$, $6.4\%$, and $6.3\%$ at the 25th, 50th, and 75th percentiles, respectively.
\end{itemize}
\noindent These findings indicate that the Top@2 MoE configuration effectively captures complementary information across diverse scene layouts, while the shared expert extracts scene-agnostic features applicable across environments. The combination of these two components results in a synergistic improvement in localization accuracy.
\begin{figure*}[t]
    \centering
    \includegraphics[width=\linewidth]{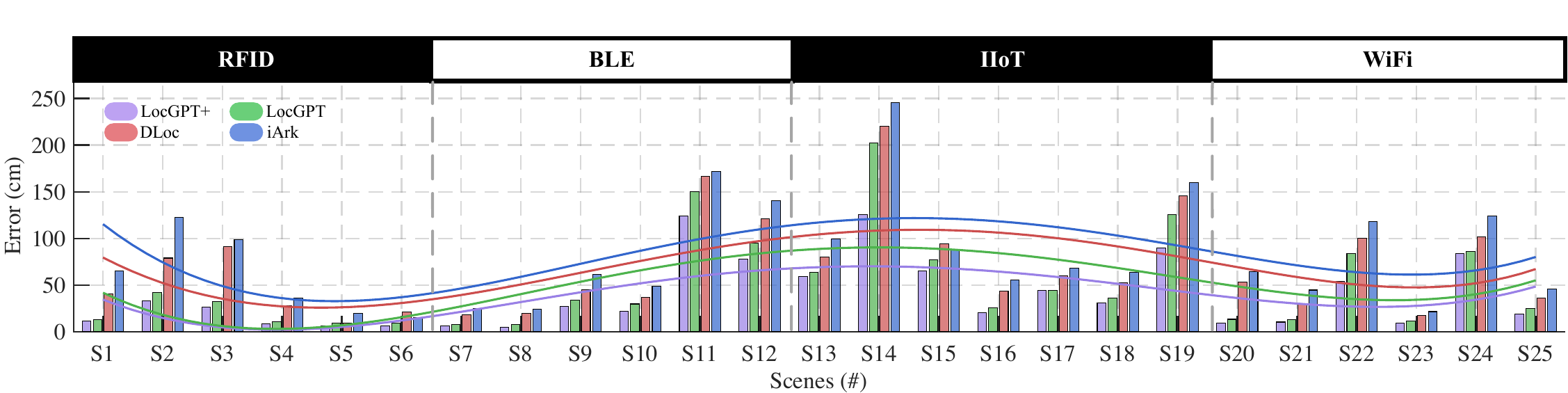}
    \caption{Overall Accuracy. \oursystem is evaluated on 25 selected test scenes, compared with LocGPT, DLoc, and iArk.}
    \label{fig:overall-exp}
\end{figure*}
\subsection{Performance of \oursystem}

Finally, we focus on evaluating the localization accuracy achieved by \oursystem. For benchmarking, we compare against three state-of-the-art deep learning-based localization models: (1) LocGPT~\cite{zhao2024understanding}, which employs a full Transformer encoder-decoder architecture; (2) DLoc~\cite{ayyalasomayajula2020deep}, which uses a ResNet-based encoder-decoder; and (3) iArk~\cite{an2020general}, which leverages a ResNet-based model to regress device locations. All models are evaluated on the 25 held-out test scenes (S1–S25). For each model, 80\% of the labeled data is used for training (DLoc and iArk) or fine-tuning (LocGPT and LocGPT+), while the remaining 20\% is used for testing. To ensure a fair comparison, all models are adapted to accept standardized inputs in the form of a $36 \times 9$ spatial spectrum.

The experimental results are shown in Fig.~\ref{fig:overall-exp}, from which we draw two key observations:
\begin{itemize}[leftmargin=*]  
  \setlength{\parskip}{0pt}
    \setlength{\itemsep}{0pt plus 1pt}
  \item First, \oursystem achieves localization errors ranging from 4.9 cm to 125 cm, with a mean error of 39.06 cm across 25 test scenes. It significantly outperforms DLoc (mean: 68.45 cm) and iArk (mean: 81.12 cm) by 42.9\% and 51.9\%, respectively. Both DLoc and iArk are trained from scratch, without any pretraining, highlighting the substantial performance gains enabled by pretraining. 
  \item Second, although LocGPT is also pretrained—using 21.3\% labeled data—it yields a higher mean error of 49.95~cm, underperforming \oursystem by 21.8\%. This performance gap can be attributed to LocGPT’s inability to digest the remaining unlabeled data, which \oursystem effectively utilizes through self-supervised pretraining.
\end{itemize}
\noindent In summary, \oursystem delivers superior localization accuracy by combining the strengths of the MoE architecture and the \rfrp pretraining framework. While the MoE improves the model’s generalization across diverse environments, \rfrp enables effective learning from large-scale unlabeled wireless data, substantially reducing the need for labeled supervision.

\subsection{Impact Analysis}

Finally, we analyze the impact of potential contributing factors on localization accuracy using fine-tuned results across the 25 test scenes.

\begin{figure*}[t]
	\centering
	\begin{minipage}[t]{0.25\linewidth} 
		\centering
		\includegraphics[width=\linewidth]{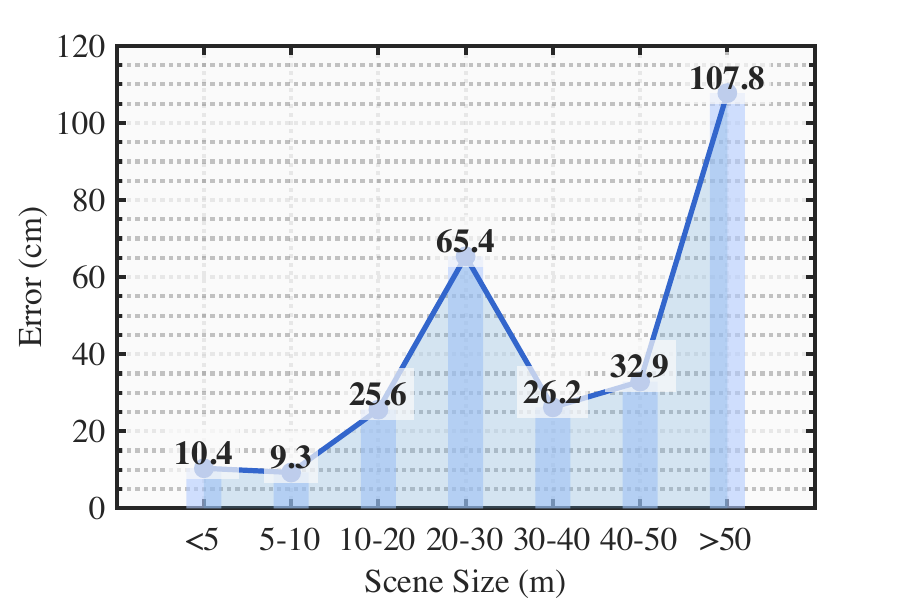}
		\caption{Impact of Scene Size}
		\label{fig:distance-exp}
	\end{minipage}%
	\begin{minipage}[t]{0.25\linewidth}
		\centering
		\includegraphics[width=\linewidth]{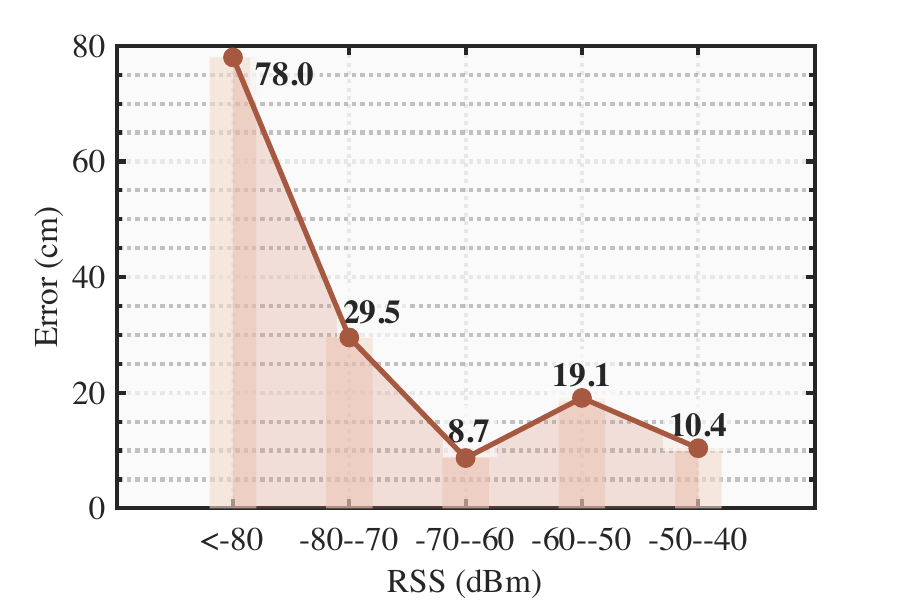}
		\caption{Impact of RSS}
		\label{fig:rss-exp}
	\end{minipage}%
	\begin{minipage}[t]{0.25\linewidth}
		\centering
		\includegraphics[width=\linewidth]{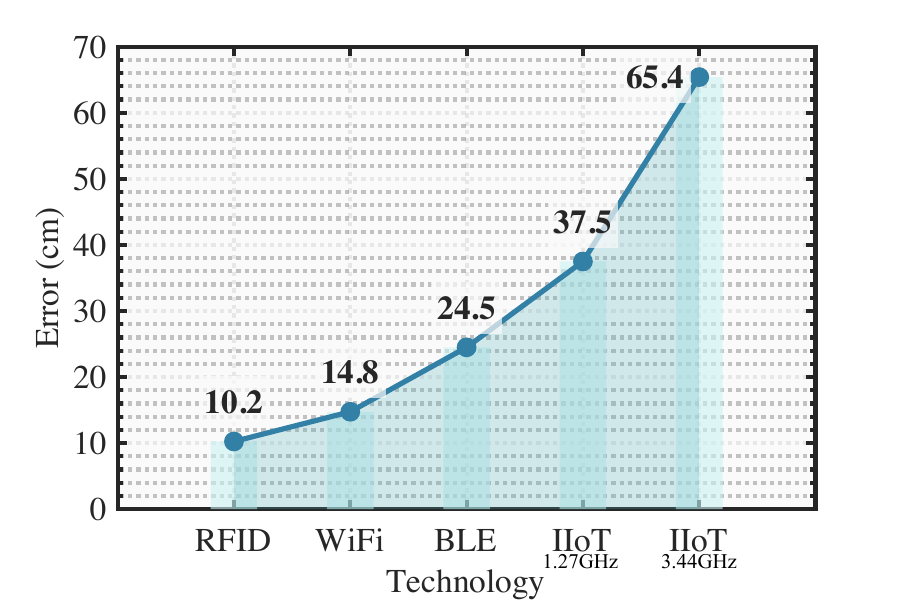}
		\caption{Impact of Technology}
		\label{fig:freq-exp}
	\end{minipage}%
	\begin{minipage}[t]{0.25\linewidth}
		\centering
		\includegraphics[width=\linewidth]{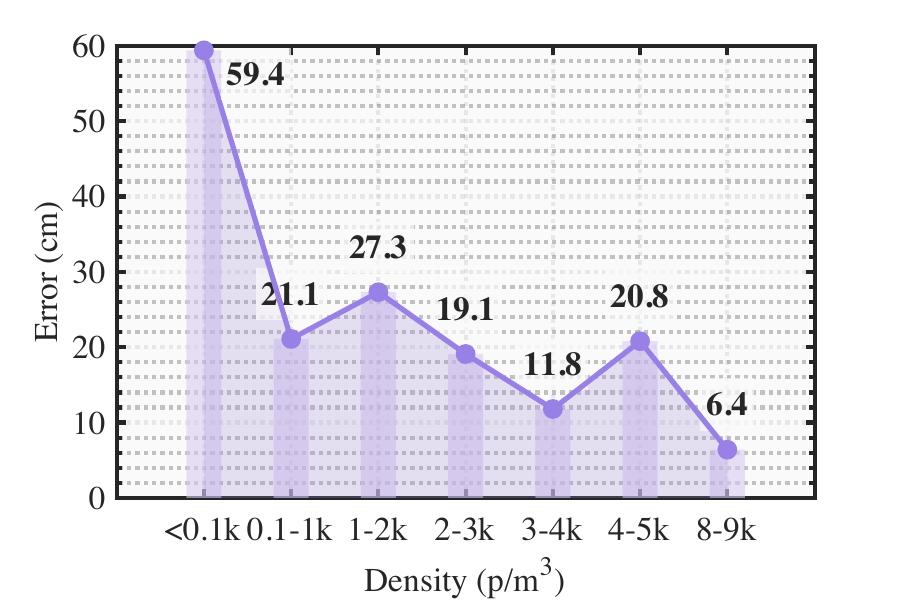}
		\caption{Impact of Density}
		\label{fig:density-exp}
	\end{minipage}
\end{figure*}

\textbf{(1) Impact of Scene Size.} The 25 tested scenes are categorized into seven size groups: $<5$ meters (extra-small), $5 \sim 10$ meters (small), $10 \sim 20$ meters (medium), $20 \sim 30$ meters (large), $30 \sim 40$ meters (ultra-large), $40 \sim 50$ meters (very large), and $>50$ meters (mega-large). The localization errors for each group are presented in Fig.~\ref{fig:distance-exp}. Overall, the localization error shows an increasing trend with scene size, primarily because greater communication distances degrade RF signal quality.
Interestingly, we observe a performance improvement (a “bounce”) in the ultra and very large groups. These measurements were taken in relatively open environments such as theaters and stadiums, where minimal obstacles reduce the impact of multipath interference.

 \textbf{(2) Impact of RSS.} The scenes are categorized into five groups based on the mean RSS: less than $-80$ dBm (very weak), $-80$ to $-70$ dBm (weak), $-70$ to $-60$ dBm (medium), $-60$ to $-50$ dBm (good), and $-50$ to $-40$ dBm (very good). The localization errors for each group are shown in Fig.~\ref{fig:rss-exp}. As expected, stronger RF signals correspond to lower localization errors, as weaker RSS values are more vulnerable to noise and NLoS effects, leading to greater signal degradation. Nonetheless, RSS alone does not fully determine localization accuracy. For example, in the $-70$ to $-60$ dBm range, the error drops to 8.7 cm—an outlier likely influenced by secondary factors such as scene layout and frequency characteristics.

\textbf{(3) Impact of Technology.} There are four wireless technologies across five frequencies in our datasets: RFID (920 MHz), WiFi (2.4 GHz), BLE (2.4 GHz), IIoT (1.27 GHz), and IIoT (3.44 GHz). We show the localization errors regarding the band in Fig.~\ref{fig:freq-exp}. The trend in localization error is clear: RFID < WiFi < BLE < IIoT (1.27 GHz) < IIoT (3.44 GHz), namely, accuracy degrades as frequency increases. A potential reason is that higher frequencies experience much more RF attenuation from obstacles, leading to the notable trends.
{\color{black}
From a signal frequency perspective, higher frequencies correspond to shorter wavelengths, leading to faster signal attenuation and increased susceptibility to absorption or reflection by obstacles (e.g., walls or human bodies). Consequently, low-frequency signals, such as RFID, exhibit more stable propagation and lower positioning errors, whereas high-frequency signals, such as IIoT at 3.44 GHz, experience rapid attenuation and higher errors. From the standpoint of signal modulation and bandwidth, RFID employs narrowband modulation, which enhances its robustness against interference. WiFi uses OFDM, which provides strong resistance to multipath interference. In contrast, BLE and IIoT signals exhibit greater variability and comparatively higher positioning errors.
}

\textbf{(4) Impact of Density.} We investigate the impact of sampling density, defined as the number of sampled positions per unit volume (samples per $m^3$). The 25 tested scenes are categorized into seven groups, covering densities from 0 to 9k samples per cubic meter (s/m³). The corresponding localization errors are shown in Fig.~\ref{fig:density-exp}. At densities below 100~s/m³, the localization error is notably high, reaching 59.4~cm. As the sampling density increases, the error generally decreases. In general, a higher sampling density provides more training data, allowing the model to better capture the signal's spatial distribution and thereby improve localization accuracy. 

\begin{figure*}[t]
    \centering
    \includegraphics[width=\linewidth]{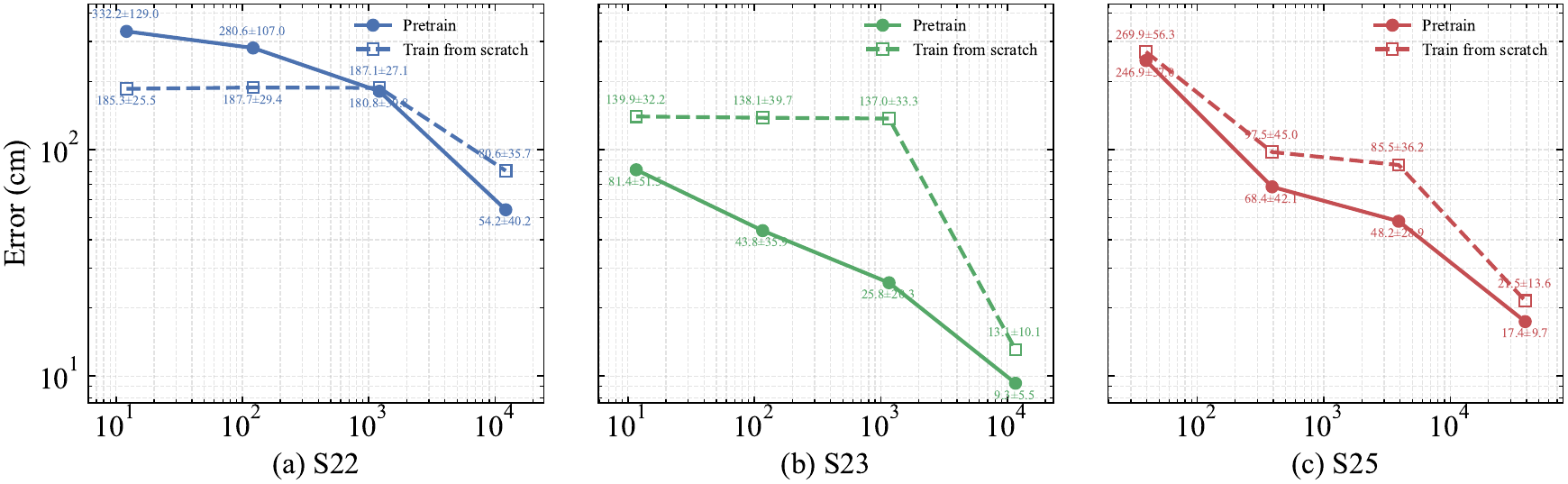}
    \caption{Sample Density Analysis. \textnormal{The x-axis represents sampling density, ranging from 10 $samples/m^2$ to over 10,000 $samples/m^2$, while the y-axis reports the median localization error.}}
    \label{fig:density}
\end{figure*}

{\color{black}
\subsection{Sample Density Analysis}
We further evaluate how localization performance varies under different sampling densities across representative scenes (S22, S23, and S25). Here, the sampling density is measured as the number of samples per unit area (samples/m$^2$). As shown in Fig.~\ref{fig:density}, reducing sampling density consistently degrades localization accuracy for all methods. However, the degradation patterns differ significantly between models trained from scratch and those pre-trained with RFGP.

At high sampling density (on the order of $10^4$ samples per unit area), pretraining provides moderate but consistent improvements (e.g., S22: 54.2\,cm vs. 80.6\,cm; S23: 9.3\,cm vs. 13.1\,cm; S25: 17.4\,cm vs. 21.5\,cm), indicating that when sufficient supervision is available, both models can achieve strong performance, with pretraining offering additional refinement.

As the sampling density decreases to approximately $10^3$ samples per unit area, the benefit of pretraining becomes substantially more pronounced. At this density, pretraining reduces the median error by up to $4\text{--}5\times$ (e.g., S23: 25.8\,cm vs. 137.0\,cm; S25: 48.2\,cm vs. 85.5\,cm). 

At even lower densities (on the order of $10^2$ and $10^1$ samples per unit area), models trained from scratch quickly saturate and fail to improve (e.g., S22 remains around $\sim$185\,cm; S23 around $\sim$140\,cm), while pretrained models degrade more gracefully and still maintain meaningful localization capability (e.g., S23: 43.8\,cm at $\sim10^2$ density; 81.4\,cm at $\sim10^1$ density).

These results highlight two key observations. First, dense sampling is beneficial but not strictly required when leveraging large-scale pretraining. Second, RFGP significantly improves data efficiency: it enables the model to retain robust performance even when labeled data is reduced by two to three orders of magnitude. This supports the practical viability of RFGP in real deployments, where dense calibration is costly and often infeasible.
}

\subsection{Latent Space Visualization and Interpretability} 
To qualitatively inspect whether pretraining changes the structure of the learned representation, we visualize the encoder features under two settings: training from scratch and fine-tuning the \rfrp-pretrained model.

\begin{figure*}[t]
\includegraphics[width=\linewidth]{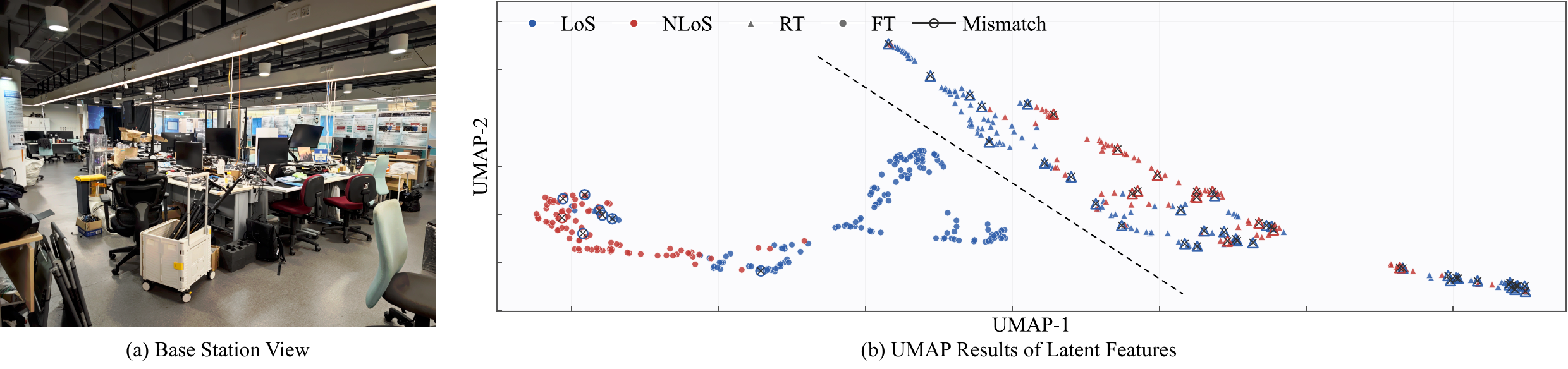}
\caption{{\color{black}Latent Space Visualization. \textnormal{(a) depicts the view of the base station in S12, while the UMAP results are shown in (b).}}}
\label{fig:vis}
\end{figure*}

 \textcolor{black}{
As illustrated in Fig.~\ref{fig:vis}, we select S12—a complex laboratory environment with severe multipath effects and heavy occlusions (e.g., walls and dense equipment)—to visualize the test-set latent features projected into a 2D space using UMAP. A sample is labeled as LoS if the direct path between the transmitter and the corresponding base station is unobstructed; otherwise, it is labeled as NLoS. It can be observed that for the model trained from scratch, the latent features of LoS and NLoS points are heavily mixed without distinct boundaries. In contrast, after fine-tuning, the pretrained model produces features that are visually more separable between LoS and NLoS samples in the UMAP projection.}

 \textcolor{black}{
Furthermore, to improve interpretability, we map severe localization errors (defined as mismatches in which a true LoS position is predicted within an NLoS region, or vice versa) to the visualization. In the fine-tuned model, severe localization errors are not uniformly distributed in the UMAP projection. Instead, they tend to appear in regions where the LoS/NLoS structure becomes ambiguous, such as near the boundary between the two groups or closer to the opposite group. For example, some true LoS samples with large errors are projected closer to the NLoS region. This pattern is consistent with the interpretation that, after \rfrp pretraining and fine-tuning, the encoder uses LoS/NLoS-related propagation cues in its representation: when severe environmental distortion makes a sample's signal pattern resemble the opposite propagation condition, the sample is also harder to localize correctly. Compared with the model trained from scratch, where LoS and NLoS features are largely mixed, this qualitative evidence suggests that \rfrp pretraining makes the latent representation more sensitive to LoS/NLoS-related structure. Since UMAP is a qualitative visualization method and can be affected by sampling choices and hyperparameters, we use this analysis as suggestive evidence rather than definitive proof of LoS/NLoS disentanglement.}

{\color{black}
\section{DISCUSSION}
\label{sec:discussion}

While the preceding sections focus on accuracy and generalization, this section discusses the practical implications of deploying RFGP in real indoor localization systems.

\subsection{Deployment Model}

RFGP is designed primarily as an offline pretraining framework rather than an end-to-end mobile localization pipeline. In our current implementation, the computationally expensive RF-NeRF decoder is used only during self-supervised pretraining to impose a physics-guided reconstruction objective. After pretraining, the decoder is not required for online localization. The deployed localization pipeline only needs the pretrained LM encoder and a lightweight prediction head to map incoming spatial spectra to device positions. This separation is important for practical deployment: the expensive physics-guided reconstruction stage can be performed offline or periodically on a server, while real-time inference can run on an edge server, an access point controller, or a building-level localization server.

This deployment model matches many indoor localization scenarios. RF spectra are typically collected by infrastructure-side devices, such as WiFi APs, BLE stations, RFID readers, or IIoT base stations. Therefore, localization inference does not necessarily need to run on the mobile or IoT device itself. Instead, the infrastructure can aggregate spectra from multiple anchors, run the localization model, and return the estimated position to the application. This design avoids imposing a heavy computational, storage, or energy burden on resource-constrained tags and mobile devices.

\subsection{Computation and Storage Cost}

The main cost of RFGP lies in offline pretraining. In our experiments, LocGPT+ contains 570M parameters and is pretrained with scene-specific RF-NeRF decoders. This design is computationally heavier than conventional localization models, but the cost is amortized across many downstream scenes and applications. Once pretrained, the same LM can be fine-tuned with less labeled data and reused across multiple environments and wireless technologies.

For practical deployments, we envision three operating modes. First, in high-accuracy infrastructure deployments such as warehouses, hospitals, factories, and large public buildings, LocGPT+ can run on an edge server connected to multiple RF anchors. Second, in medium-resource settings, the pretrained model can be compressed through quantization, pruning, or expert selection, since only a small subset of experts may be frequently activated in a given building. Third, for mobile-only deployments, LocGPT+ can serve as a teacher model to distill a smaller student model. The student model would sacrifice some accuracy but would be better suited to on-device inference. We leave a systematic study of compression and distillation as future work.

\subsection{Data Collection and Sampling Density in Practice}

Another practical concern is the amount and density of data required by RFGP. Our large-scale dataset uses dense spatial sampling to support controlled evaluation and to study cross-scene generalization. However, dense-labeled sampling is not required for the intended deployment workflow. RFGP is motivated by the observation that unlabeled RF signals can be collected continuously and at low cost from existing wireless infrastructure. In a real deployment, RF spectra can be accumulated passively from normal device traffic, periodic beaconing, or opportunistic user movement. Only a smaller set of labeled points is needed for downstream calibration and fine-tuning. 

The sampling-density results suggest that denser data generally improves localization performance, but they should not be interpreted as requiring exhaustive 3D labeling in every new scene. Instead, there is a trade-off between deployment effort and localization accuracy. For applications requiring centimeter-level accuracy, such as robotic navigation or industrial asset tracking, a denser calibration phase may be justified. For coarse-grained indoor navigation or context-aware services, a lower sampling density may be sufficient. Future deployments can also combine active learning with uncertainty estimation to decide where additional labels are most valuable, thereby reducing unnecessary data collection.

\subsection{Limitations and Future Directions}

RFGP still has several limitations. First, the current model is large, and its pretraining cost is significant. While this cost is manageable for offline foundation-model-style training, it may be too high for small-scale deployments. Promising strategies to reduce inference and storage costs include model compression, expert pruning, and teacher-student distillation.

Second, the current discussions about deployment assume that infrastructure-level sensing relies on known anchor locations. In scenarios involving fully ad-hoc deployments, rapidly changing environments, or poorly calibrated anchors, additional adaptation mechanisms may be necessary. Incorporating online calibration, continual learning, or uncertainty-aware model updates could enhance RFGP's robustness for long-term real-world operations.

Another limitation is that our current evaluation primarily measures the effectiveness of the RF-NeRF decoder through downstream localization and decoder ablations. We do not directly verify whether the learned attenuation or radiance fields accurately reflect true material properties or specific physical propagation parameters. Future work should focus on comparing learned RF fields with controlled channel measurements, calibrated ray-tracing simulations, or material-level annotations.

Lastly, our evaluation emphasizes localization accuracy and generalization, but real deployments also need to consider factors like latency, energy consumption, privacy, and maintenance costs. Conducting a comprehensive end-to-end deployment study across various edge platforms is a crucial direction for future research.

}

\section{CONCLUSION}
\label{section:conclusion}

 \textcolor{black}{
In this paper, we presented RFGP, a novel self-supervised framework designed to overcome the cross-scene generalization bottleneck in RF-based indoor localization. Moving beyond standard pretraining approaches, RFGP uniquely couples a large localization encoder with a physics-informed RF-NeRF decoder. This asymmetric architecture encourages the model to learn transferable spatial-spectral representations from unlabeled RF spectra. Extensive evaluations on a massive-scale dataset—comprising over 7.3 million samples across 100 diverse scenes and four wireless technologies—demonstrate that RFGP significantly outperforms both train-from-scratch and standard supervised pretraining baselines. Ultimately, by successfully embedding physical propagation priors into the self-supervised learning loop, RFGP substantially reduces the reliance on costly manual annotation, paving the way for highly scalable, generalizable, and cost-effective indoor localization models.}

\begin{acks}
We thank all the anonymous reviewers for their valuable comments and helpful suggestions.
This work is supported by UGC/GRF (No. 15211024), the Innovation and Technology Fund (ITS/034/23FP, GHP/079/22SZ), and the Shenzhen Fundamental Research Scheme (No. JCYJ20230807140410022). This work is also supported by the Research Institute for Artificial Intelligence of Things (RIAIoT).
\end{acks}

\newpage
\bibliographystyle{ACM-Reference-Format}
\bibliography{locgpt,iark}

\end{document}